\documentclass[twoside,twocolumn,9pt]{article}
\usepackage{extsizes}
\usepackage[super,sort&compress,comma]{natbib} 
\usepackage[version=3]{mhchem}
\usepackage[left=1.5cm, right=1.5cm, top=1.785cm, bottom=2.0cm]{geometry}
\usepackage{balance}
\usepackage{mathptmx}
\usepackage{sectsty}
\usepackage{graphicx} 
\usepackage{lastpage}
\usepackage[format=plain,justification=justified,singlelinecheck=false,font={stretch=1.125,small,sf},labelfont=bf,labelsep=space]{caption}
\usepackage{float}
\usepackage{fancyhdr}
\usepackage{fnpos}
\usepackage[english]{babel}
\usepackage{array}
\usepackage{droidsans}
\usepackage{charter}
\usepackage[T1]{fontenc}
\usepackage[usenames,dvipsnames]{xcolor}
\usepackage{setspace}
\usepackage[compact]{titlesec}
\usepackage{hyperref}
\usepackage{amsmath}

\usepackage{epstopdf}

\definecolor{cream}{RGB}{222,217,201}

\def\ci{\citep}
\def\cia{\citep}

\def\eq#1{eq.~\eqref{#1}}
\def\beq{\begin{equation}}
\def\eeq{\end{equation}}
\def\Cpl{C_{\rm Lo+GM1}^{\rm planar}}
\def\Cvesi{C_{\rm Lo+GM1}^{\rm vesicle}}
\def\kLo{\kappa_{\rm Lo}}
\def\kLd{\kappa_{\rm Ld}}

\begin{document}

\pagestyle{fancy}
\thispagestyle{plain}
\fancypagestyle{plain}{
\renewcommand{\headrulewidth}{0pt}
}

\makeFNbottom
\makeatletter
\renewcommand\LARGE{\@setfontsize\LARGE{15pt}{17}}
\renewcommand\Large{\@setfontsize\Large{12pt}{14}}
\renewcommand\large{\@setfontsize\large{10pt}{12}}
\renewcommand\footnotesize{\@setfontsize\footnotesize{7pt}{10}}
\renewcommand\scriptsize{\@setfontsize\scriptsize{7pt}{7}}
\makeatother

\renewcommand{\thefootnote}{\fnsymbol{footnote}}
\renewcommand\footnoterule{\vspace*{1pt}%
\color{cream}\hrule width 3.5in height 0.4pt \color{black} \vspace*{5pt}} 
\setcounter{secnumdepth}{5}

\makeatletter 
\renewcommand\@biblabel[1]{#1}            
\renewcommand\@makefntext[1]%
{\noindent\makebox[0pt][r]{\@thefnmark\,}#1}
\makeatother 
\renewcommand{\figurename}{\small{Fig.}~}
\sectionfont{\sffamily\Large}
\subsectionfont{\normalsize}
\subsubsectionfont{\bf}
\setstretch{1.125} 
\setlength{\skip\footins}{0.8cm}
\setlength{\footnotesep}{0.25cm}
\setlength{\jot}{10pt}
\titlespacing*{\section}{0pt}{4pt}{4pt}
\titlespacing*{\subsection}{0pt}{15pt}{1pt}

\fancyfoot{}
\fancyfoot[LO,RE]{\vspace{-7.1pt}\includegraphics[height=9pt]{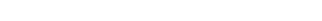}}
\fancyfoot[CO]{\vspace{-7.1pt}\hspace{13.2cm}\includegraphics{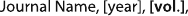}}
\fancyfoot[CE]{\vspace{-7.2pt}\hspace{-14.2cm}\includegraphics{head_foot/RF}}
\fancyfoot[RO]{\footnotesize{\sffamily{\thepage}}}
\fancyfoot[LE]{\footnotesize{\sffamily{\thepage}}}
\fancyhead{}
\renewcommand{\headrulewidth}{0pt} 
\renewcommand{\footrulewidth}{0pt}
\setlength{\arrayrulewidth}{1pt}
\setlength{\columnsep}{6.5mm}
\setlength\bibsep{1pt}

\makeatletter 
\newlength{\figrulesep} 
\setlength{\figrulesep}{0.5\textfloatsep} 

\newcommand{\topfigrule}{\vspace*{-1pt}%
\noindent{\color{cream}\rule[-\figrulesep]{\columnwidth}{1.5pt}} }

\newcommand{\botfigrule}{\vspace*{-2pt}%
\noindent{\color{cream}\rule[\figrulesep]{\columnwidth}{1.5pt}} }

\newcommand{\dblfigrule}{\vspace*{-1pt}%
\noindent{\color{cream}\rule[-\figrulesep]{\textwidth}{1.5pt}} }

\makeatother

\twocolumn[
  \begin{@twocolumnfalse}
\vspace{1em}
\sffamily
\begin{tabular}{m{4.5cm} p{13.5cm} }
& \noindent\LARGE{\textbf{There and back again: bridging meso- and nanoscales to understand lipid vesicle patterning$^\dag$}} \\
 & \vspace{0.3cm} \\

 & \noindent\large{Julie Cornet,$^a$ Nelly Coulonges,$^{a,b}$ Weria Pezeshkian,$^c$ Ma\"el Penissat-Mahaut,$^b$ Siewert-Jan Marrink,$^d$ Nicolas Destainville, $^{a,\ddag}$ Matthieu Chavent,$^{b,e,\mbox{\scriptsize \textsection}}$ and Manoel Manghi,$^{a,\P}$} \\



\end{tabular}

 \end{@twocolumnfalse} \vspace{0.6cm}

  ]

\renewcommand*\rmdefault{bch}\normalfont\upshape
\rmfamily
\section*{}
\vspace{-1cm}


\footnotetext{\textit{$^{a}$~Laboratoire de Physique Théorique, Université de Toulouse, CNRS, UPS, France.}} 
\footnotetext{\textit{$^{b}$~Institut de Pharmacologie et Biologie Structurale, Université de Toulouse, CNRS, Université Toulouse III – Paul Sabatier, 31400, Toulouse, France. }}
\footnotetext{\textit{$^{c}$~Niels Bohr International Academy, Niels Bohr Institute, University of Copenhagen, Blegdamsvej 17, 2100 Copenhagen, Denmark.}}
\footnotetext{\textit{$^{d}$~Groningen Biomolecular Sciences and Biotechnology Institute, University of Groningen, Nijenborgh 7, 9747 AG Groningen, The Netherlands.}}
\footnotetext{\textit{$^{e}$~Laboratoire de Microbiologie et Génétique Moléculaires (LMGM), Centre de Biologie Intégrative (CBI), Université de Toulouse, CNRS, UPS, France.}}


\footnotetext{\dag~Electronic Supplementary Information (ESI) available at end: supplementary text and 6 figures.}

\footnotetext{\ddag~E-mail: nicolas.destainville@univ-tlse3.fr}
\footnotetext{\textsection~E-mail: matthieu.chavent@univ-tlse3.fr}
\footnotetext{$\P$~E-mail: manoel.manghi@univ-tlse3.fr}




\sffamily{\textbf{We describe a complete methodology to bridge the scales between nanoscale Molecular Dynamics and (micrometer) mesoscale Monte Carlo simulations in lipid membranes and vesicles undergoing phase separation, in which curving molecular species are furthermore embedded. To go from the molecular to the mesoscale, we notably appeal to physical renormalization arguments enabling us to rigorously infer the mesoscale interaction parameters from its molecular counterpart. We also explain how to deal with the physical timescales at stake at the mesoscale. Simulating the so-obtained mesoscale system enables us to equilibrate the long wavelengths of the vesicles of interest, up to the vesicle size. Conversely, we then backmap from the meso- to the nano- scale, which enables us to equilibrate in turn the short wavelengths down to the molecular length-scales. By applying our approach to the specific situation of the patterning of a vesicle membrane, we show that macroscopic membranes can thus be equilibrated at all length-scales in achievable computational time offering an original strategy to address the fundamental challenge of time scale in simulations of large bio-membrane systems.
}}\\

\rmfamily 


\section{Introduction}

\begin{figure*}
\centering
\includegraphics[width=\textwidth]{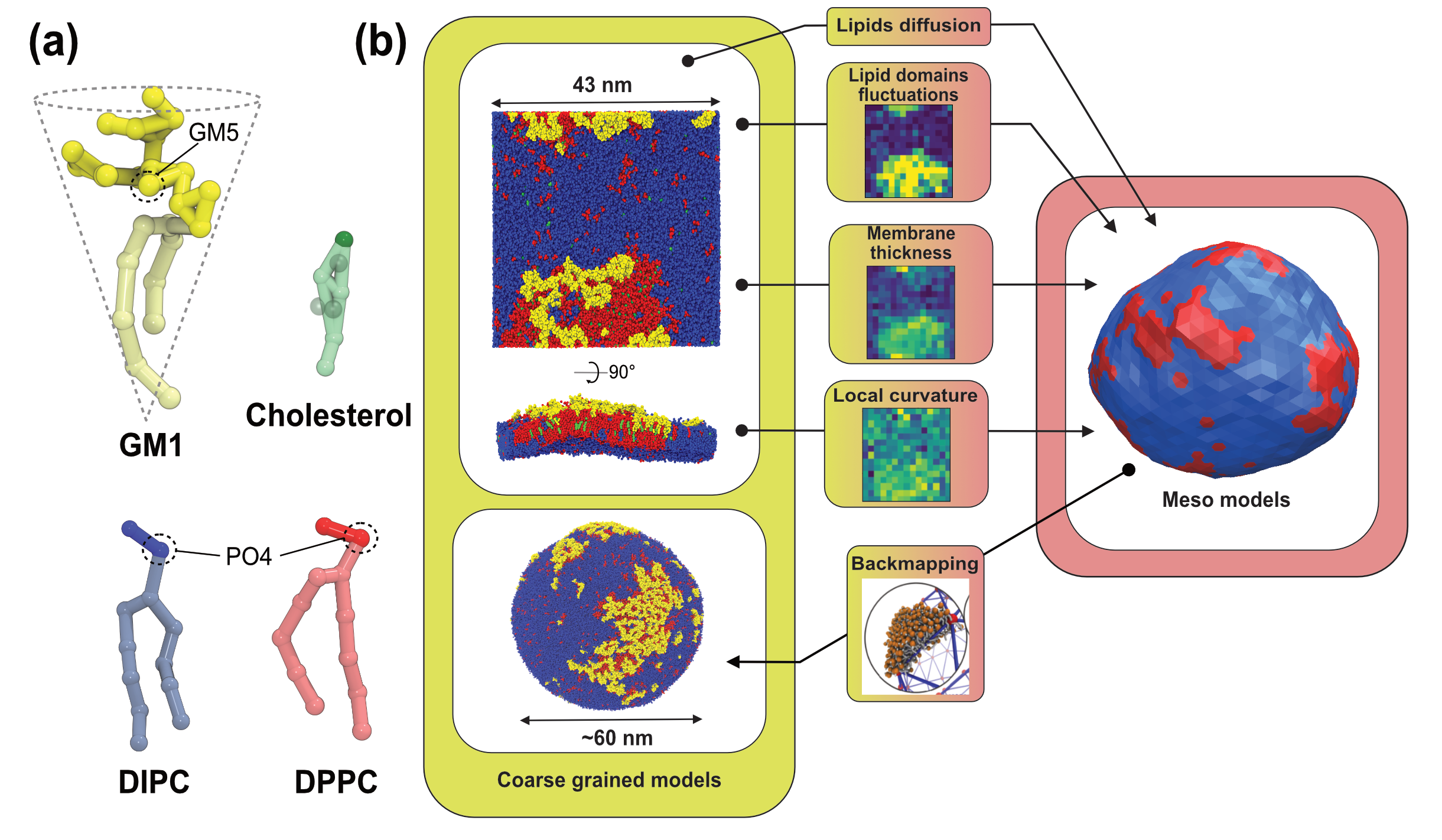} 
\caption{\textbf{Principle of our multiscale modeling scheme of lipid membranes.} \textbf{(a)} Lipid models used in CG simulations. GM1 model corresponding to a monosialotetrahexosylganglioside, cholesterol model (with virtual sites in transparency \citep{Melo2015}), DIPC model corresponding to 1,2-dilino\-leoyl-sn-gly\-cero-3-phosphocholine, DPPC model corresponding to 1,2-dipal\-mitoyl-sn-glycero-3-phosphocholine. The most hydrophilic parts of the lipids are represented in darker colors while the lighter colors depict the more hydrophobic parts of the models. The PO4 and GM5 beads used to define the membrane surface are outlined on the different lipids (see Method section). \textbf{(b)} From CG molecular dynamics simulations (upper-left panel), we infer physical parameters required by the Mesoscale Monte Carlo model (right panel) consisting of a bi-phasic tessellated vesicle. The extraction of these parameters is illustrated in the central panels. They are: (i) two dynamical parameters relating the time-scales of lipid diffusion in the membrane plane, and transverse membrane fluctuations; (ii) lipid domain boundary fluctuations from which is inferred the line tension between both lipid phases; (iii) the membrane thickness difference between phases, from which bending moduli ratio can be inferred, and (iv) the spontaneous curvature of the Lo phase induced by insertion of GM1 in the upper leaflet. Conversely, the Meso-to-CG backmapping at the full vesicle scale (lower-left) is illustrated in the lower-middle panel. }
\label{fig:workflow}
\end{figure*}

Physics of living systems is permanently confronted to the multiplicity of length- and time- scales of interest: from the nanoscopic molecular scale where events occur below the nano-second time\-scale, to the micrometric cellular scale where microseconds or seconds are at stake. Different physical and numerical techniques have been developed to study these different scales, which are either quantum or classical, with explicit or implicit water, including stochasticity or hydrodynamics. But bridging these scales is still a difficult task as methods employed at each scale may not use the same physical dynamics~-- e.g., Molecular Dynamics (MD) or Monte Carlo (MC)~-- nor the same integration scheme~-- e.g., Newtonian or Brownian dynamics. Thus even if recently multi-scale workflows are emerging~\citep{Pezeshkian2019,Ingolfsson2022,Ingolfsson2023}, \textcolor{black}{this type of approaches has been limited in term of spatial/timing range they can span~\citep{Vickery2021} and they has been generally constrained to "one spatial direction"~\citep{Newport2018,Peng2023,Wassenaar2014}, from the nanoscale to the meso/micro-scale or the other way around.} Here, thanks to recent advances in the field~\citep{Pezeshkian2020,Pezeshkian2021}, we detail and rationalize a method to fully bridge, forward and backward, nano- and meso- scales in biomembranes (see Figure~\ref{fig:workflow}). 

We develop this approach in the specific context of lipidic biomembrane patterning. With the recent \emph{in vivo}, \emph{in vitro}, and \emph{in silico} developments, it is now recognized that cell membrane components are not homogeneously distributed, but are organized into functional lipid and protein sub-micrometric domains ~\citep{Veatch2023,Veatch2008_critical,Lang2010,Ingolfsson2014,Sezgin2017,Cebecauer2018,Duncan2020}. These nanodomains are playing fundamental roles in cell biology, especially as they serve as platforms or micro-reactors for many biological functions such as infections (viral or bacterial), cell adhesion, transport of solutes, or signaling. Thus, deciphering the formation and evolution of these domains is essential to fully understand these fundamental biological processes. Among the mechanisms proposed for their formation, the role of spontaneous curvature induced by different membrane constituents, with specific shapes, is the focus of a particular attention~\citep{review}. 

For example, the glycosphingolipid GM1 has a bulky head comprised of four monosaccharides conferring to this lipid a globally conical shape~\citep{Marsh2010,Cebecauer2018} (see Figure~\ref{fig:workflow}-a). Present predominantly in the outer cell membrane~\citep{Dasgupta2018,Liu2019}, this lipid can act as a membrane anchor for different toxins, bacteria and viruses~\citep{Ewers2010, Aureli2016} and play important roles in several neuronal processes and diseases, notably auto-immune ones~\citep{Schengrund2015, Aureli2016}. Although knowledge about its lateral organization remains an active field of research~\citep{Liu2019}, the properties of GM1 are thought to be linked to its propensity to curve the membrane, ensuing from its conical shape~\citep{Sonnino1994,Dasgupta2018,Shimobayashi2016}. To test this hypothesis on well-defined model systems, Giant Unilamellar Vesicles (GUVs) were used to better characterize {\em in vitro} the relationship between curvature generation and lipid domain formation \citep{Bao2011,Shimobayashi2016,Dasgupta2018,Steinkuhler2020}. Yet, fully linking microscopy visualization, with an intrinsically limited resolution, to nanoscale partitioning of the lipids, is still a challenge. 

On the one hand, MD simulations are the suitable tool to understand how GM1 perturbs the lipid membrane organization at the molecular scale, notably by curving the bilayer~\citep{Sreekumari2018}, or by destabilizing lipid-lipid phase separation in multicomponent membranes~\citep{Liu2019}. On the other hand, describing the effect of phase separation combined with intrinsic curvature at vesicular or cellular scales makes the use of a much larger degree of coarse-graining inevitable. Here, we illustrate how to bridge the theoretical gap between Coarse-Grained Molecular Dynamics (CG-MD) simulations and \textcolor{black}{mesoscale (Meso for short) Monte Carlo simulations \citep{Duncan2023,Ruhoff2023} } focusing on vesicular systems, as illustrated in Figure~\ref{fig:workflow}. Using  physically relevant concepts, we first show which data to extract at the nanoscale, obtained with CG-MD simulations, and how they can be used to parametrize the mesoscale Monte Carlo model. Going from nano- to mesoscales amount to integrate out microscopic degrees of freedom, which raises nontrivial so-called renormalization issues~\citep{Feigenson2014}, that we fully take into account in this work. Then we show how this leads to meaningful results for the Mesoscale model, allowing one to equilibrate membrane patterning at the scale of a whole vesicle, and how this model can predict experimental data at micrometric scales. Finally, we explain how to scale back to CG models to equilibrate in their turn the short length-scales, and decipher the overall organization of large lipid domains down to the nanoscale. We also discuss the feasibility of our approach and the limitations inherent to each model when one passes from one scale to another. This thorough approach to link the scales combined with careful explanations of how to extract meaningful data will pave the way for further development of fully integrated multiscale workflows.

\section{Methods}
\label{mat:meth}

\subsection{Coarse-grained (MARTINI) molecular dynamics simulations}
\label{sec:MARTINI}

We use the CHARMM-GUI \textit{MARTINI Maker}~\citep{Hsu2017,Qi2015} to generate $43\times43$~nm$^2$ lipid bilayer systems. The membrane patches are surrounded by about 4 nm of water on each side. Ions are added to neutralize the system.  We study a mixture of  C16:0 dipalmitoyl PC (DPPC), C18:2 dilinoleoyl PC (DLiPC), also named DIPC in the MARTINI force field, and cholesterol with concentration ratios 30:58:12 and with varying concentration of C(d18:1/18:0) N-stearoyl-D-erythro (GM1) in the upper leaflet (given in mol/mol) (see Table~\ref{tab:CG_system}). 

The CG-MD simulations are performed using the MARTINI v2.2~\citep{Marrink2007} force field in the NPT ensemble and run with GROMACS 2016 software~\citep{Abraham2015}. The temperature is set to 310~K where Lo and Ld phases coexist for these lipid mixtures. We use the velocity rescale thermostat~\citep{Bussi2007} coupled to a semi-isotropic Parinello-Rahman barostat~\citep{Parrinello1981} with pressure of 1 bar. The standard time step of 20 fs is used for all simulations. All systems are equilibrated following equilibration steps as described in the Membrane Builder workflow~\citep{Wu2014}. For production, all the planar systems were simulated for 20 $\mu$s to ensure convergence. 
 
The analysis scripts were written in Python3 using MDAnalysis packages~\citep{Gowers2016}. Especially, we use the LeafletFinder tool that allows  identifying upper and lower leaflets. 

\begin{table}
\begin{center}
\begin{tabular}{lcc}
\hline
System           & Particles & Duration \\ \hline
DPPC-DIPC-CHOL    & 167,409 & 20 \\ \hline
DPPC-DIPC-CHOL + 2.5\% GM1    & 279,234 & 20 \\ \hline 
DPPC-DIPC-CHOL + 5\% GM1      & 181,285 & 20 \\ \hline
DPPC-DIPC-CHOL + 7.5\% GM1      & 279,981 & 20 \\ \hline
DPPC-DIPC-CHOL + 10\% GM1      & 181,756 & 20 \\ \hline
DPPC-DIPC-CHOL + 15\% GM1      & 207,647 & 20 \\ \hline
Vesicle   & 3,487,094 & 10 \\ \hline
\end{tabular}
\end{center}
\caption{Summary of CG systems simulated (see also Figure~\ref{fig:GM1}).The vesicle has composition (DPPC-DIPC-CHOL + 15\% GM1). Durations are in $\mu$s.
\label{tab:CG_system}}
\end{table}

\subsection{Mesoscale model}

The Mesoscale model used in the present work has been developed and validated previously in one of our groups. We give an overview of the model in this section, the interested reader can refer to previously published work for further details~\citep{GG2,Cornet2020}.

\subsubsection{Discretization of the membrane model}
\label{sec:dis}
We discretize space, time, and accordingly the calculation of the system free energy. We consider the initial vesicle as a tessellated sphere composed of $N$ vertices~\ci{GG2}. An initial icosahedron is tessellated iteratively. This leads to accessible values for the total number of vertices $N=10 \times 4^k+2$ \ci{GG1}. Each vertex represents a small patch of one of the two species (delineated to the Vorono\"i cell associated to the vertex). The real size of this patch depends on the average vesicle radius $R$. In the case of weak shape deformations, the area $A_0$  of a patch is approximately $A_0\simeq4\pi R^2/N$. For example, for $N=2562$ sites ($k=4$ iterations), in a small vesicle of radius $R=100$~nm, a patch would contain about $100$ lipids. Two neighbor vertices are separated by the average lattice spacing 
\begin{equation}
a = \left( \frac{8 \pi}{\sqrt 3}\right)^{\frac12} \frac{R}{\sqrt N}.
\label{eq:a}
\end{equation}

\subsubsection{Helfrich free energy}
\label{sec:dis_Helfrich}
In the continuous case, the Helfrich elastic free energy is
\beq
H_{\rm Helf}=\frac12\int \kappa (2H-C)^2 {\rm d}S + \sigma \mathcal{A}
\eeq
where $\kappa$ is the (local) bending elastic modulus, $\sigma$ the surface tension and $\mathcal{A}$ the total vesicle area. $2H$ is the total curvature, \textit{i.e.} the sum of the two principal curvatures, and $C$ is the (local) spontaneous curvature imposed by the molecular species. We do not account for the Gaussian curvature in this model. Using the Laplace-Beltrami operator~\ci{Meyer} for the curvature term, \textcolor{black}{the discrete elastic free energy} reads
\beq
H_{\rm Helf}= \frac{1}{2}\sum_i \kappa_i [2H_i - C_i]^2 \mathcal{A}_i + \sigma \sum_i \mathcal{A}_i
\label{eq:discrete:Helfrich}
\eeq
with $\mathcal{A}_i$ the area associated to a vertex ($\mathcal{A}_i=A_0$ at start). The total curvature $2H_i$ is obtained as the signed norm of the Laplace-Beltrami operator $\mathbf{K}_i$  
\beq
\mathbf{K}_i=\frac{1}{2\mathcal{A}_i}\sum_j(\cot \alpha_{ij} + \cot \beta_{ij})(\mathbf{r}_i-\mathbf{r}_j)
\label{eq:LB}
\eeq
Here $\mathbf{r}_i$ is the position of vertex $i$ and the sum is taken over its first neighbors $j$ and the angles $\alpha_{ij}$ and $\beta_{ij}$ are the angles of the two triangles sharing the edge $(\mathbf{r}_i ,\mathbf{r}_j)$ and opposite to this edge \citep{GG2,Cornet2020}. \textcolor{black}{Both $\kappa_i$ and $C_i$ might be vertex-dependent in the case of bi- or multi-phasic systems, see below.} We use the reduced membrane tension $\tilde \sigma =  \sigma R^2 /(k_{\rm B} T)$. 

The vesicle volume $V$ is fixed close to the the initial volume $V_0$ by a hard quadratic constraint. Hence the following energy term is added to the Helfrich energy
\beq
H_{\rm V}=\frac{1}{2}K_v\left(\frac{V}{V_0}-1\right)^2
\eeq
with $K_v=2\times10^6 k_{\rm B}T$. In contrast, the total vesicle area is constrained by a soft constraint and controlled by the surface tension $\sigma$. 

\subsubsection{Species characteristics and interaction}
\label{sec:species}

We use the discrete two-dimensional Ising (or lattice-gas) model to describe the Lo/Ld binary mixture~\ci{Chaikin}. The fact that membrane lipid binary mixtures belong to the 2D Ising universality class has been experimentally checked in Ref.~\ci{Veatch2008_critical}. The Ising model Hamiltonian reads  
\beq
H_I= -J_I\sum_{\langle i,j \rangle} s_i s_j 
\label{eq:Ising}
\eeq
where $s_i = \pm 1$ and the sum runs over nearest-neighbor vertex pairs. The tessellation of the sphere that we use is a triangular lattice, each vertex having 6 neighbors, except 12 vertices originating from the original icosahedron, which have only 5 neighbors. The lattice-gas composition, $\phi_i =0$ or 1, on any vertex $i$ of the tessellated lattice is related to its ``spin'' $s_i$ through $\phi_i =(1+s_i)/2 $. The Ising constant $J_I>0$ measures the strength of short-range affinities between membrane constituents. In our case, we work at fixed $\sum_i s_i$ and thus at fixed concentration $\phi=\sum_i \phi_i/N$ and temperature $T$. Varying the temperature between the different simulations of the Ising model amounts to varying the Ising coupling $J_I$. In our simulations we chose to fix the temperature $T$ and to tune the value of $J_I$. Its exact critical value where the phase transition takes place is $J_{I,c}= (\ln3/4)k_{\rm B}T\simeq k_{\rm B}T/3.64$ on an infinite triangular lattice~\ci{Baxter}, which means that if $J_I<J_{I,c}$ (resp. $J_I>J_{I,c}$), then the system is in the disordered (resp. ordered) phase. 

The local spontaneous curvature $C_i=C(\phi_i)$ entering Eq.~\eqref{eq:discrete:Helfrich} depends on the local concentration $\phi_i$ and can take two values in the mescoscopic model, $C_0=2/R$ for the Ld phase and $\Cvesi$ for the Lo+GM1 one (see below). Similarly, the local bending modulus $\kappa_i=\kappa(\phi_i)$ can take two values, $\kappa_{\rm Ld}$ and $\kappa_{\rm Lo}$. In this work, for simplicity, $\kappa_{\rm Lo}$ does not depend on the GM1 concentration in the simulations.

We used the Mesoscale code developed in Ref.~\cite{GG2,Cornet2020} to simulate fluctuating vesicles. At each Monte Carlo step, two local move attempts are applied to randomly chosen vertices: (1)~a single vertex undergoes a small radial displacement $\delta r$ according to Metropolis' rule, which locally modifies the elastic energy; (2)~the compositions $\phi_i$ of two neighbor vertices are swapped, modifying the interaction energy as well as the elastic energy through the elastic parameters $\kappa(\phi_i)$ and $C(\phi_i)$ above. Since we consider a system with conserved order parameter $\phi$, we use the Kawasaki algorithm~\ci{Barkema}. 

Contrary to alternative models in the spirit of the Dynamic Triangulated Surface (DTS) one~\citep{Hu2011,Feigenson2014,Penic2015,Pezeshkian2019,Siggel2022,PezeshkianIpsen2023}, edge flips are not allowed in the tessellated system \textcolor{black}{and vertex moves are only radial}, consequently no constraint on edge lengths needs to be applied in the present model. In compensation, our model is valid in the vesicle quasi-spherical regime only. For more information, the whole simulation scheme is described and discussed in detail in Ref.~\ci{Cornet2020}.  

\subsection{Extraction of line tension from domain boundary fluctuations}
\label{sec:line:tension}

\begin{figure}
\centering
\hspace{-0.5cm}
\includegraphics[width=5cm]{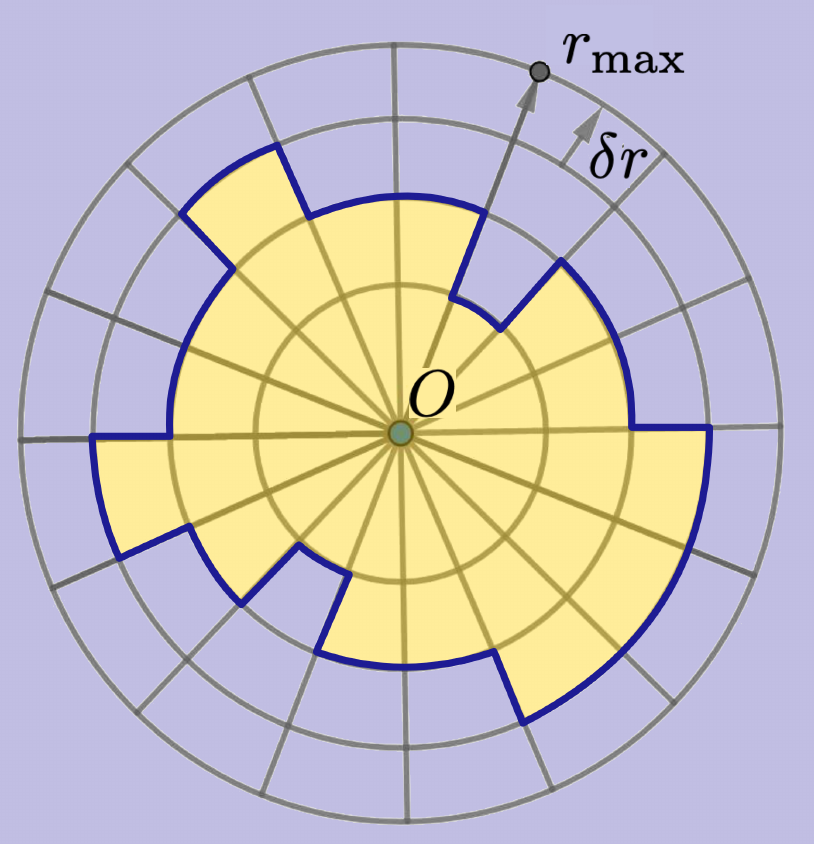} 
\caption{Principle of the polar discretization of the Lo domain in order to locate its boundary (dark blue line) with a (discrete) polar function $r(\theta)$. The ``dartboard'' is divided in $n_c=r_{\rm max}/\delta r$ concentric circles and $n_a$ regularly spaced angular sectors, thus defining $n_c \times n_a$ boxes in which the majority rule is applied to determine whether they belong to the Lo (yellow) or the Ld (light blue) phase. }
\label{fig:target}
\end{figure}

As schematized in Figure~\ref{fig:target}, we use a new discretization of the plane, in the form of a ``dartboard'' made of $n_c$ concentric circles, centered on the domain geometric barycenter and separated by $\delta r=2$~nm, as well as $n_a=20$ angular sectors. In each box of the so-obtained ``dartboard'', we use the same majority rule as in the square mesh (see Section~\ref{identLoLd} below) so as to identify the boxes belonging to the Lo domain and those belonging to the Ld surrounding phase. Starting from the center and following the rays, we then identify the boundary when at least two successive Ld boxes (or the last circle) have been encountered, so that small bubbles of the Ld phase are not considered as being part of the boundary. This defines a discrete version of the domain boundary $r(\theta)$. 

We then use Eq. (3) of Ref.~\ci{Esposito2007} to directly link the power spectrum of the boundary fluctuations to the line tension $\lambda$ of the mixture. However, 
we use a different convention for the Fourier coefficients:
\begin{equation}
u_n = \frac1{2\pi} \int_0^{2\pi} u(\theta) e^{-ni\theta} {\rm d}\theta
\end{equation}
that we estimate by discretizing the integral. Here $u(\theta)$ is defined by $r(\theta)=R_0[1+u(\theta)]$. Then it can be shown that
\begin{equation}
\langle|u_n|^2\rangle = \frac{k_{\rm B}T}{2 \lambda \pi R_0(n^2-1)}
\label{eq:lambda}
\end{equation}
in equilibrium, for $n>1$. The mode $n=1$ corresponds to a translation of the whole domain and its value is non-physical as it simply reflects the error possibly made when identifying the domain barycenter. In this equation $R_0$ is the equivalent radius of a non-fluctuating domain of area $\pi R_0^2$. In principle, it is different from the average domain radius $\langle r(\theta)\rangle=R_0\left(1-\sum_{n > 0} \langle|u_n|^2\rangle \right) <R_0$~\ci{Esposito2007}. In practice, however, $R_0$ is close to  $\langle r(\theta)\rangle$ for the line tension values at play here and we identify both.

\subsection{Relation between the CG extracted parameters and the Mesoscale model ones}
\label{sec:link:scales}

In order to bridge the two modeling scales, the coarse-grained and mesoscale ones, we need to establish the link between the parameters that we measure in the CG-MD simulations with the input parameters of the mesoscale description. This can lead to technical difficulties exposed below. 

Owing to renormalization issues \ci{Feigenson2014}, the Ising parameter depends on the coarse-graining level because it must account for the microscopic degrees of freedom integrated out in the coarse-graining process. We denote by 
\begin{itemize}
\item[$\bullet$] $a$ the simulation lattice spacing at the mesoscale, \textit{i.e.} the average length-size of the elementary tessellation triangles, see Eq.~\eqref{eq:a}.
\item[$\bullet$] $l_0\sim 1$~nm the typical distance between lipids at the molecular scale,
\item[$\bullet$] $J_{I,0}$ the Ising interaction parameter at the molecular scale,  of  critical value  $J_{I,c}$. It is directly related to the Flory parameter $\chi$ of the binary mixture~\citep{Flory}.
\end{itemize}
We assume that the interaction network at the molecular scale can be assimilated to a triangular lattice, owing to the symmetries of bidimensional liquids.
Close to the critical point, one has
\beq
J_I-J_{I,c}=\frac{a}{l_0}\left(J_{I,0}-J_{I,c}\right)
\label{eq:J:renorm}
\eeq
thus relating the two scales~\ci{Feigenson2014}. On this lattice $k_{\rm B}(T_c-T)=(4/\ln3)\left(J_{I,0}-J_{I,c}\right)$ close to $T_c$, where $T_c$ can be measured either in experiments or in molecular dynamics simulations.
After measuring the line tension $\lambda$ as explained in the text, we can have access to the Ising parameter at the molecular scale. Indeed, close to the critical point, $\lambda$ depends algebraically on $J_{I,0}-J_{I,c}$ and vanishes at the critical point. Renormalization arguments also show that both quantities are in fact proportional for the 2D Ising model universality class. More precisely, 
\beq
\lambda = \frac{4 \sqrt{3}}{l_0}(J_{I,0}-J_{I,c}).
\label{eq:link1}
\eeq
close to $J_{I,c}$. The prefactor depends on the lattice and is equal to $4 \sqrt{3}$ on a triangular lattice~\cia{Shneidman2001}.
Owing to \eq{eq:J:renorm}, we thus also have 
\beq
\lambda=\frac{4 \sqrt{3}}{a}(J_I-J_{I,c}).
\label{eq:link2}
\eeq
This enables us to relate the Ising parameter of the mesoscale simulations and the line tension with the lattice spacing $a$ through \eq{eq:link2}, measured in the CG simulations. 

The analysis of bilayer simulations at the molecular level thus provides realistic membrane parameter values that can be injected into the Mesoscale model. In this way, the simulations can then be tuned in order to tackle different biologically relevant systems at large length and time scales. Note that the bending moduli $\kappa$ depend only logarithmically on the scale $a$~\ci{Safran}, and we thus consider them as constant for sake of simplicity. 

\subsection{Backmapping from the mesoscale to the coarse-grained scale}
\label{sec:backmapping}

The Mesoscale model can be backmapped to CG resolution (Martini 2.2) based on the TS2CG approach of Ref.~\cite{Pezeshkian2020}. We follow the different steps presented in the TS2CG tutorial (http://cgmartini.nl/index.php/2021-martini-online-workshop/tutorials/558-9-ts2cg). In brief, the last frame from the Mesoscale model is saved in a dynamic triangulated surface (DTS) file format readable by the TS2CG program. In this file, both Lo and Ld domains were defines as well as the overall shape of the vesicle using labeled triangles. TS2CG inputs such as lipids concentrations and Area Per Lipid (APL) are extracted from the previous CG simulation of a planar system with 15\% of GM1. These values are presented in Table~\ref{tab:Meso_input}. 

\begin{figure*}[t]
\centering
\includegraphics[width=16cm]{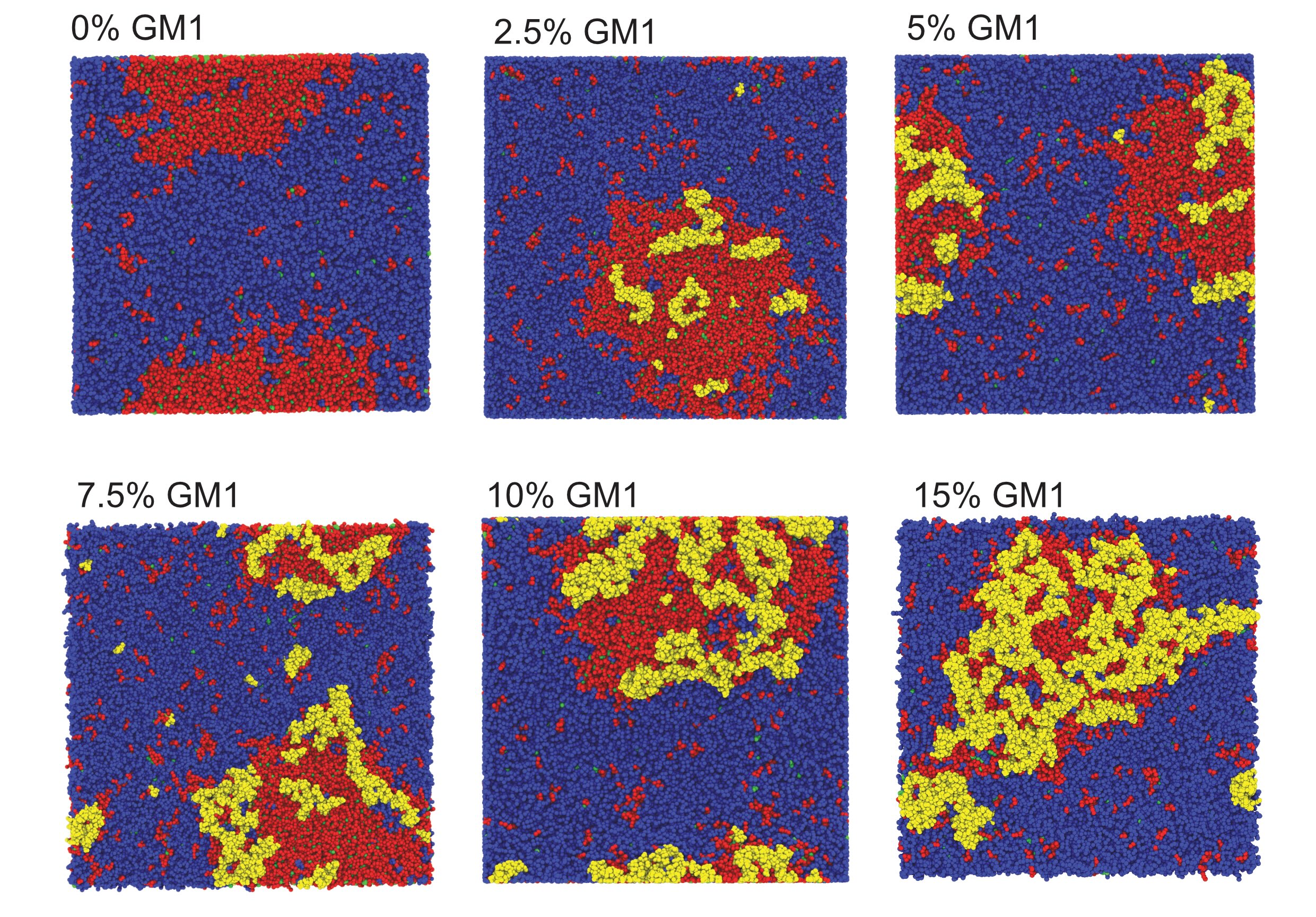} 
\caption{{Top view of CG-MD membrane systems with different molar concentration of GM1 at $20~\mu$s. The membrane phase separates in all the systems forming a Lo domain enriched in DPPC, GM1, and cholesterol surrounded by a Ld domain enriched in DIPC. Due to boundary conditions, the Lo domain is sometimes visually divided in two parts but it is a visualization artefact. GM1 lipids are depicted in yellow, DPPC in red, DIPC in blue, and cholesterol in green. }}
\label{fig:GM1}
\end{figure*}

\begin{table}
\begin{center}
\begin{tabular}{lccc}
\hline
Lipid type  & Upper leaflet & Lower leaflet & APL \\ \hline
 \multicolumn{4}{c}{Ld Domain} \\ \hline
DPPC & 0.06 & 0.06 & 0.68 \\ \hline
DIPC & 0.91 & 0.91 & 0.77 \\ \hline 
CHOL & 0.03 & 0.03 & 0.5  \\ \hline
 \multicolumn{4}{c}{Lo Domain} \\ \hline
DPPC & 0.40 & 0.68 & 0.68 \\ \hline
DIPC & 0.04 & 0.04 & 0.77  \\ \hline 
CHOL & 0.28 & 0.28 & 0.5 \\ \hline
GM1 & 0.28 & 0 & 0.7 \\ \hline

\end{tabular}
\end{center}
\caption{Input parameters used by TS2CG (see also Figure~\ref{fig:GM1}). APL means  Area Per Lipid, in nm$^2$.
\label{tab:Meso_input}}
\end{table}
 
We perform 1000 steps of standard energy minimization and 5000 steps of molecular dynamics run with lipids headgroup position restrained to relax the lipid chains. We perform these steps without solvent particles using the Dry Martini force field~\citep{Arnarez2015}. Next, the vesicle membrane is solvated by propagating an equilibrated Martini water box in the system and removing any water particle within a certain cutoff from the membrane particles as done previously~\citep{Pezeshkian2023str}. 

The solvated system is then equilibrated following several steps of equilibration as done for the planar systems (see above). For these steps, the constraints beads, called Wall, are present to keep the overall shape of the vesicle (see~\citep{Pezeshkian2020} for more details). Then, we remove the wall beads and performed 10~$\mu$s of production using the same thermostat and barostat parameters than the planar systems (see above) with an isotropic pressure coupling. 

In equilibrated vesicles, both leaflets do not have the same area for geometrical reasons and thus have different molecule numbers. In  experiments on vesicles, equilibration is a rather slow process ensured by thermally activated lipid flip-flops. Such flip-flops  cannot occur at the rapid simulation timescale, so that one must take care of filling in each leaflet with the proper molecule number. This is guaranteed by TS2CG.

\section{Results}

\subsection{Nanoscale analysis of lipid phase separation and GM1 partitioning}

 We have performed CG-MD simulations of $43\times43$ nm$^2$ bilayers made of a DPPC-DIPC-Chol mixture with varying molar concentrations of GM1 molecules in the upper leaflet from 0\% up to 15\% (see Methods section).  The duration of these simulations is set to $20~\mu$s. The DPPC-DIPC-Chol mixture allows these systems to undergo a phase separation where coexist a phase enriched in DPPC and Cholesterol molecules called liquid-ordered (Lo) and another phase enriched in DIPC,   the liquid-disordered phase (Ld) ~\citep{Jefferys2014, Risselada2008, Pantelopulos2018, Liu2019}, as displayed in Figure~\ref{fig:GM1}.

\subsubsection{\textit{Convergence to equilibrium depends on GM1 concentration}}

\begin{figure}[t]
\centering
\includegraphics[width=8cm]{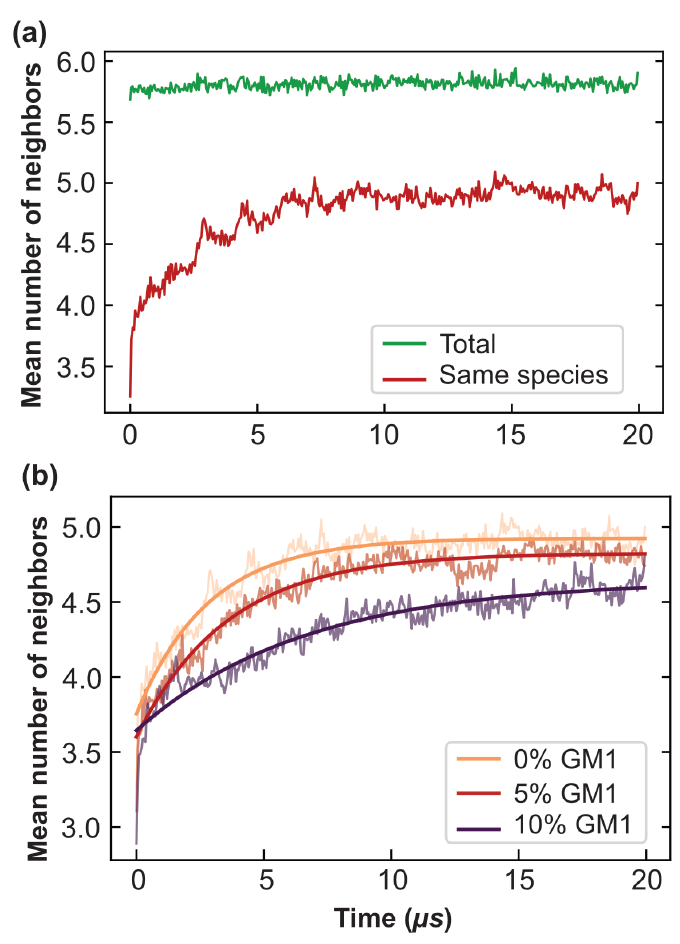}
\centering
\caption{Mean number of neighbors and convergence to equilibrium in function of time. (a) Mean number of neighbors of the same species for each lipid over simulation time in DPPC-DIPC-Chol (30:58:12) mixture, without GM1 (red). In green is represented the total number of neighbors, whatever their nature. The cut-off distance $d=$13\;{\AA} was used to determine the neighbors of each lipid. (b) Mean number of neighbors of the same species fitted with $\langle n(t) \rangle =A-B e^{-t/\tau}$. This fit allows one to extract the typical equilibration time $\tau$. One gets $\tau \simeq$ 2.8, 3.6 and 6.6~$\mu$s for GM1 fractions in the upper leaflet of 0 (orange, same as red curve in a), 5\% (red) and 10\% (purple) respectively.}
\label{fig:neigh}
\end{figure}

In order to quantify phase separation degree and its evolution with time, we compute the mean number of neighbors of the same species for each lipid throughout the simulation and plot it over time. Indeed, in a binary mixture controlled by short range interactions, the average energy per molecule is directly directed to the average composition of its immediate neighborhood. 
We use a cut-off distance $d=$13\;{\AA} to define the neighborhood zone, in order to have about six neighbors per molecule in equilibrium, as in a simple two-dimensional liquid. Without GM1, this observable reaches a plateau shortly after $5~\mu$s (Figure~\ref{fig:neigh}-a). Its value is lower than the average number of neighbors since a finite fraction of molecules have some neighbors of a different species. To compare systems with different concentrations of GM1, we fit this curve with a single exponential $\langle n(t) \rangle =A-B e^{-t/\tau}$ as shown in Fig.~\ref{fig:neigh}-b. We measure relaxation times $\tau$ on the order of a few $\mu$s. This relaxation time increases with the concentration of GM1, suggesting that the presence of GM1 slows down the phase separation process as observed previously~\citep{Liu2019}. Thus, these results \textcolor{black}{suggest} that our systems can phase separate in between $5~\mu$s and $10~\mu$s in function of the GM1 concentration. 

\subsubsection{\textit{Identification of Lo and Ld phases}}
\label{identLoLd}

\begin{figure}
\centering
\includegraphics[width=9cm]{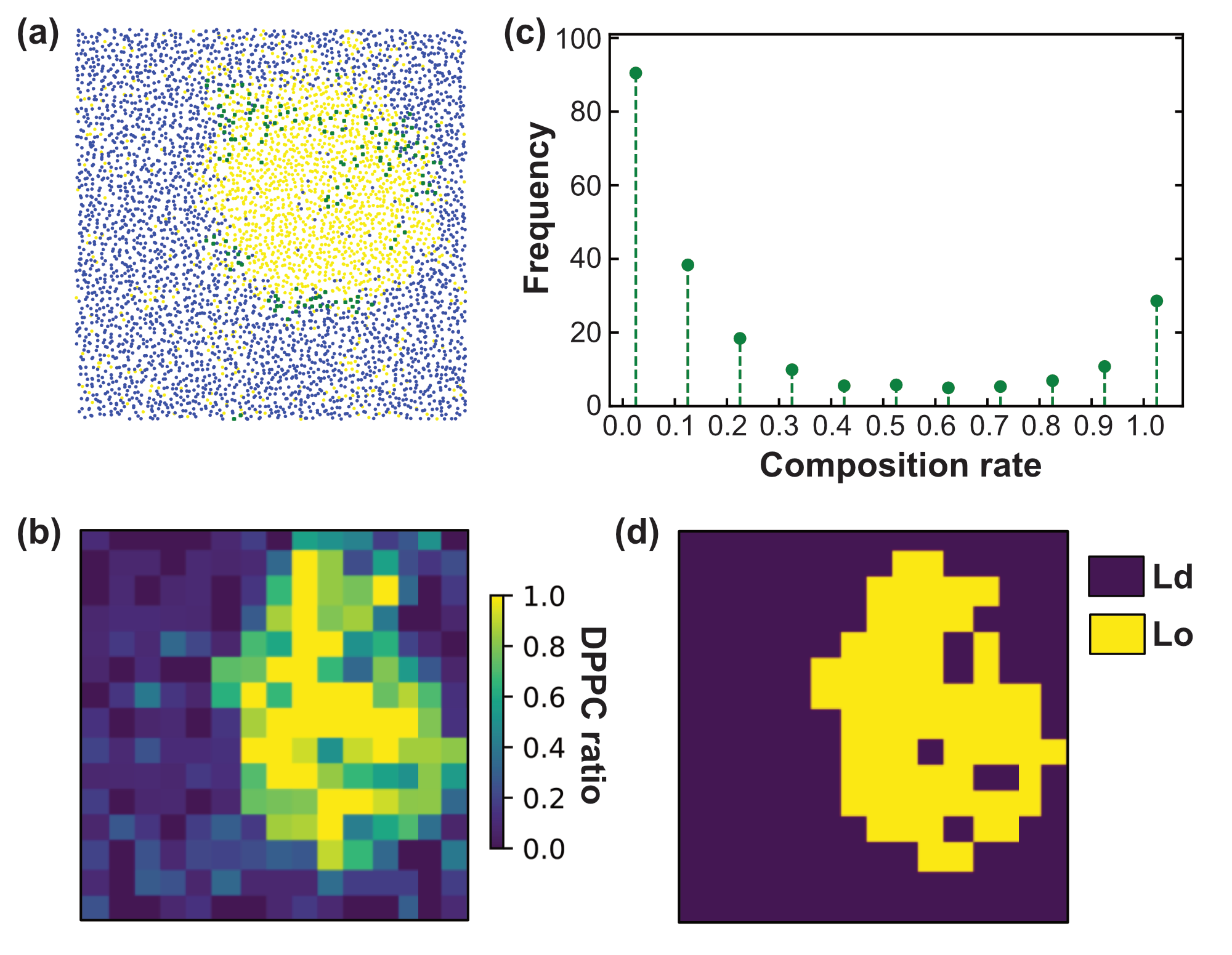} 
\caption{\textbf{Lo/Ld discretisation of the CG membrane model. (a)} Illustration of the spatial distribution of the different lipid species for the last frame of planar CG system (see Figure~\ref{fig:GM1}). Each molecule position is identified by the position of one bead in the leaflet plane (see text). DPPC in yellow, DIPC in blue and GM1 in green. \textbf{(b)} Examples of analysis performed for a DPPC-DIPC-Chol (30:58:12) mixture with 10\% GM1 in the upper leaflet (using a different shapshot as compared to (a)). The bilayer is divided into a 15$\times$15 mesh. \textbf{(c)} Composition distribution (fraction of DPPC in boxes of the bilayer, measured throughout simulation time). \textbf{(d)} The composition repartition matrix in terms of DPPC ratio is binarized in Lo (yellow) and Ld (blue) boxes based on the composition distribution presented in c and according to the majority rule described in the text.}
\label{fig:compo_dist}
\end{figure}

Once-phase-separated, we can  identify the different phases by discretizing the membrane patches into a mesh of $N=L\times L$ boxes with $L=15$ (each box is then of side $\delta x\simeq 3$~nm and contains at least one lipid). For DIPC and POPC lipids, each phospholipid position is identified by the coordinates of the PO4 bead corresponding to the phosphate group while GM1 positions is identified by the coordinates of the GM5 bead (Figure~\ref{fig:workflow}-a). The latter is located at an average depth similar to the phosphate groups in the upper leaflet as the two first sugar rings of GM1 can be deeply embedded in the interfacial region of the membrane ~\citep{Liu2019}. The local composition of a box can then be computed as the ratio of any molecules to the total number of molecules in the box (Figure~\ref{fig:compo_dist}-a,b). For DPPC molecules, the fraction of molecules in all boxes is sampled over all simulation frames and the threshold was set to a DPPC fraction of 0.6, where the composition distribution shows a marked minimum (Figure~\ref{fig:compo_dist}-c). The composition ratio is then regularly computed in each box along the course of the simulation and binarized with the threshold defined above so that the two phases are identified (Figure~\ref{fig:compo_dist}-d). We can now compute observables of interest in function of the local composition in the mesh and measure their correlations (see ESI$^\dag$).

\subsubsection{\textit{Localization of GM1}}

\begin{figure}[t]
\centering
\includegraphics[width=8cm]{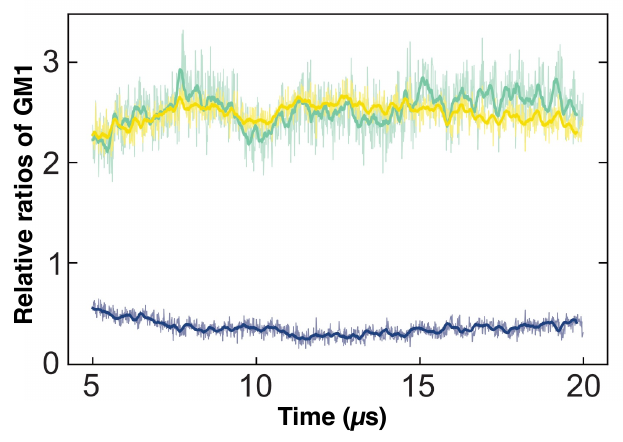} 
\caption{{Relative ratios of GM1 molecules located in the Ld (blue) and Lo (yellow) phases and at the Lo phase boundary (green) weighted by the corresponding area ratios. DPPC-DIPC-Chol (30:58:12) mixture with 10\% GM1 in the upper leaflet.}}
\label{fig:ratios}
\end{figure}

To begin with, we measure the relative ratios of GM1 molecules that are located in the Lo and Ld phases and at the Lo phase boundary, weighted by the corresponding area ratios (Figure~\ref{fig:ratios}). We compute for instance the number of GM1 molecules in the Lo phase over the Lo domain area and divide it by the total number of GM1 molecules over the leaflet area. This confirms that in our simulations, the GM1 molecules preferentially partition into the Lo phase as observed experimentally in Ref.~\cite{Shimobayashi2016}. We conduct the same measurement for the GM1 molecules located in the Lo domain boundary zone in order to see whether GM1 preferentially get located at the boundary. Our results indicate that they are roughly homogeneously distributed in the domain (see also Figure~\ref{fig:GM1}). Thereafter, the ensuing curvature is thus assumed to be quasi-uniform in the domain.
 
\subsection{Calculation of parameters extracted from CG-MD simulations to accurately design the Mesoscale model}

These first calculations allow us to calculate parameters extracted from CG simulations to be used as inputs for the Mesoscale model (see  Methods section). They are: 
\begin{enumerate}
  \item the bending moduli of both phases ($\kLd$ and $\kLo$),
    \item the two dimensionless spontaneous curvatures of the Ld and Lo+GM1 phases, denoted by $RC_0$ and $R\Cvesi$,
    \item the Ising parameter $J_I$ (related to the line tension between both phases), 
    \item the relative timescales associated with diffusion in the membrane plane on the one hand and transverse deformation modes on the other hand, if one is interested in dynamical properties.
\end{enumerate}
We explain below how to extract from CG simulations the values of these parameters, which are intrinsic to the molecular species at play (and their phase state through temperature). As the system needs at most around $8~\mu$s to equilibrate (see Figure~\ref{fig:neigh}-b), the following measurements are performed after $8~\mu$s, for a duration of $12~\mu$s (unless stated otherwise) so that phase separation is essentially reached.

Two other parameters are important for membrane patterning~\citep{Weitz2013,review,Cornet2020}: 
\begin{itemize}
  \item the membrane surface tension $\sigma$,
  \item the area fraction of each phase ($\phi$ is the fraction of the Lo+GM1 phase and $1-\phi$ the fraction of the Ld phase, both conserved through time)
\end{itemize}
These extrinsic parameters depend on the experimental conditions and are not extracted from the CG simulations.

\subsubsection{\textit{Determination of bending moduli of both phases $\kLd$ and $\kLo$}} 

Both Ld and Lo phases bending modulus values, respectively $\kLd$ and $\kLo$, must be set as parameters in the Mesoscale model. 

Generically, Ref.~\cite{Fowler2016} has shown how the bending modulus $\kappa$ of a bilayer made of a homogeneous lipid mixture can be inferred from the spectral density of the membrane thermal shape fluctuations in CG simulations using the Helfrich model for membranes. 
In our CG simulations, the surface tension was set to $\sigma=0$, to make easier the determination of $\kappa$. However, the continuous Helfrich model ignores short-wavelength molecular scales where individual lipids jiggle out of the local membrane plane, forming some ``protrusions''~\citep{Goetz1999,Imparato2005,Schmid2013}. Assuming that protrusions and Helfrich fluctuations are independent random variables, one gets the spectral density of fluctuations in the Fourier space~\citep{Shiba2011} 
\begin{equation}
\langle | \hat h({\mathbf q}) |^2 \rangle = L^2 \, k_{\rm B} T \left(\frac{1}{\kappa q^4} +  \frac{1}{\sigma_{\rm pr} q^2} \right)
\label{h2q}
\end{equation}
where $\sigma_{\rm pr}$ is the protrusion tension, measured to be  $\sigma_{\rm pr} \sim 0.1$~J/m$^2$ on simple coarse-grained numerical membrane models~\citep{Goetz1999,Imparato2005}. Protrusions dominate the  spectral density of fluctuations for wavevectors $q \geq \sqrt{\sigma_{\rm pr} /\kappa}$, \textit{i.e.} at wavelengths $2\pi/q \leq 4$~nm
when $\kappa \sim 10 k_{\rm B} T$. Thus one must use large enough system sizes $L$ to reliably measure $\kappa$, as well as long simulation times to reduce statistical errors.

We performed such measurements on our own CG simulations of Ld and Lo membranes, by using Ref.~\cite{Fowler2016} approach. Fitting with Eq.~\eqref{h2q}, we estimated the values of $\kappa$ for both systems, $\kLd = 13  k_{\rm B}T$ for the Ld phase, and $\kLo = 25 k_{\rm B}T$ for the Lo one (see the ESI$^\dag$ for further details).

\subsubsection{\textit{Determination of the spontaneous curvature of the Lo phase}}

In order to compute the local curvature of the upper leaflet for the different phases of the CG model, \textcolor{black}{as the membrane exhibits only small bending, 
we approximate the total curvature by} a 2D (discrete) Laplacian $\tilde\Delta$, using the four nearest neighbors, applied on the height field:
\begin{equation}
\tilde\Delta h_{k,\ell} = \frac{h_{k+1,\ell}+h_{k-1,\ell}+h_{k,\ell+1}+h_{k,\ell-1}-4 h_{k,\ell}}{\delta x^2}
\label{Laplacian}
\end{equation}
Here $h_{k,\ell}$ is the height of the membrane on the box of integer coordinates $(k,\ell)$ on the $15 \times 15$ mesh defined above, endowed with periodic boundary conditions. Note that we choose to measure the curvature of the upper leaflet \textcolor{black}{because we are primarily interested by the curvature generated by GM1 insertions in this leaflet.}

\begin{figure}
\centering
\includegraphics[width=8cm]{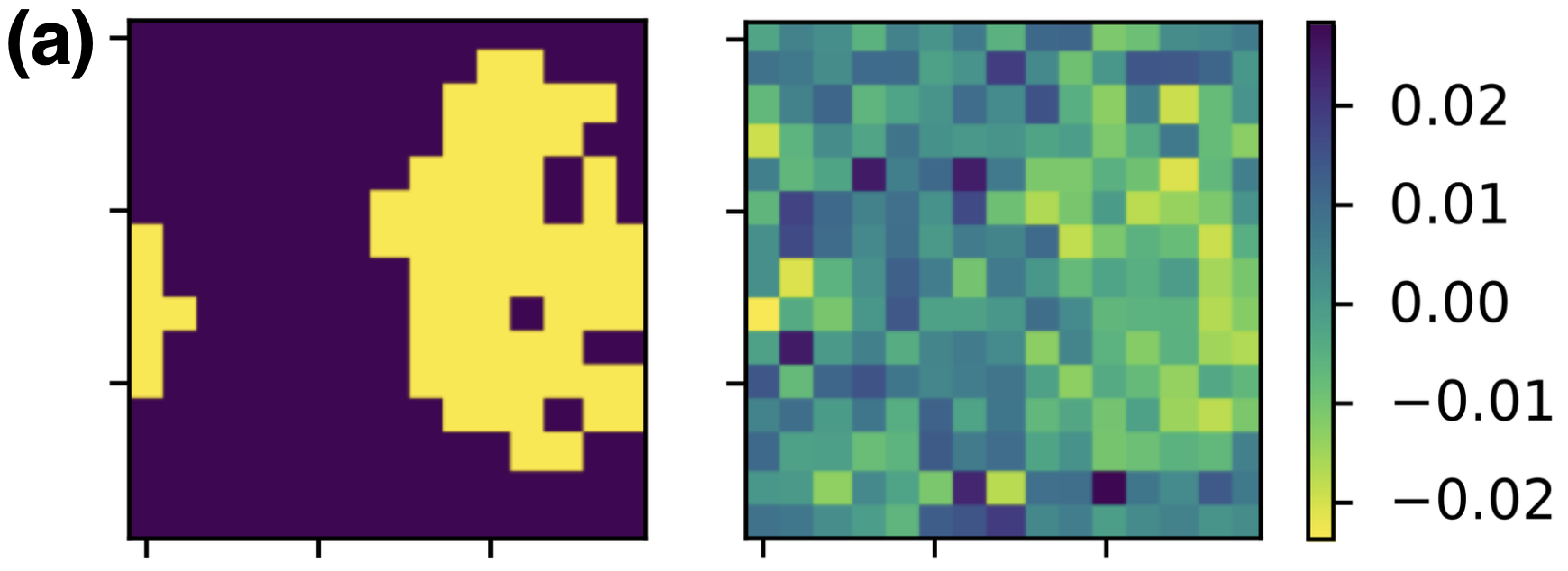} \\
\includegraphics[width=8cm]{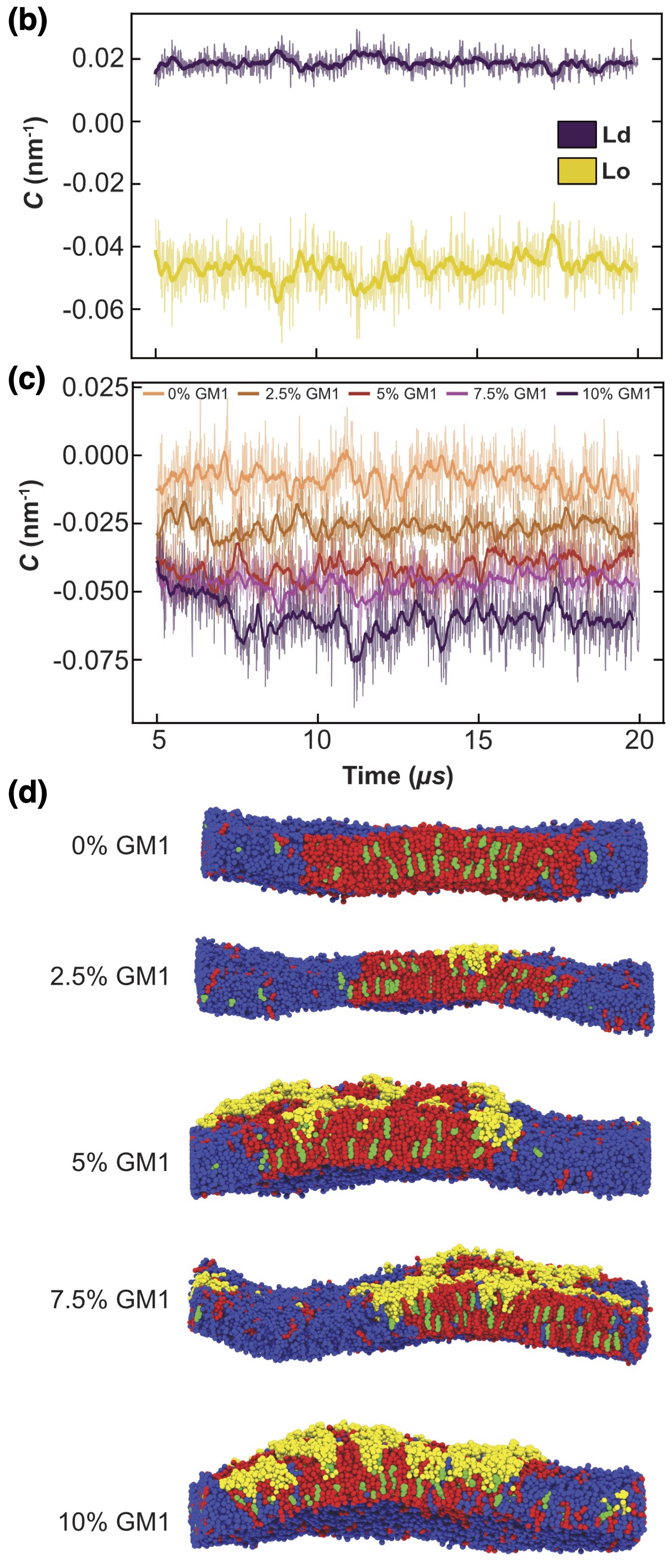} 
\caption{ Local curvature imposed by GM1 molecules. (a)~Local curvature (right, in nm$^{-1}$) against lipid phase (left). (b)~Local curvature measurement through time, averaged for Lo and Ld phase separately, in the upper leaflet for a DPPC-DIPC-Chol mixture with 7.5\% GM1 in the upper leaflet. (c)~Curvature of the Lo phase from 0 to 10\% GM1 in the upper leaflet (by increments of 2.5) from orange to purple. This curvature is negative because of our sign convention in Equation~\eqref{Laplacian}. (d) Side views of the different systems at $20~\mu$s.  }
\label{fig:curv}
\end{figure}

The Lo phase, where GM1 are mainly inserted, as already shown in Figure~\ref{fig:ratios}, displays a higher curvature than the Ld one (Figure~\ref{fig:curv}-a,b). This Lo domain curvature increases with the addition of GM1 molecules (Figure~\ref{fig:curv}-c). We can notice that for 0\% of GM1, the Lo domain already has a weak curvature. This upper leaflet weak curvature results from the difference of thickness in between the Lo and Ld domains and the fact that the leaflet position is identified through the positions of the PO4 and GM5 beads (see above).

Note that because of periodic boundary conditions, the total curvature averaged on the whole system mathematically vanishes \textcolor{black}{when it is approximated by a Laplacian}. Thus if the Lo phase is curved, the Ld phase must be curved as well in the opposite direction, so that the average curvature is zero (Figure~\ref{fig:curv}-d). 

\begin{figure}[t]
\centering
\includegraphics[width=9cm]{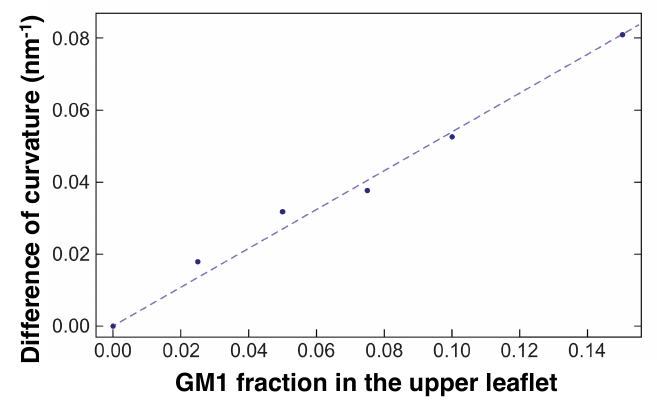} 
\caption{Correlation between curvature and GM1 concentration. Difference of curvature $\Cpl$ between the Lo domain with GM1 and the reference Lo domain without GM1 in planar geometry. The curvature values are averaged on the last 10~$\mu$s. DPPC-DIPC-Chol (30:58:12) mixture with growing fraction of GM1 in the upper leaflet, and linear fit given in Eq.~\eqref{eq:C1} (dashed line). 
\label{fig:C_GM1}}
\end{figure}

To quantify how GM1 molecules affect upper leaflet curvature, we plot (the absolute value of) the difference of curvature $\Cpl$ between the Lo domains enriched in GM1 and the small reference curvature ($|C|\simeq0.0087$~nm$^{-1}$) of Lo domain without GM1 insertions (Figure~\ref{fig:C_GM1}). The curvature values are averaged from 10 to 20~$\mu$s on the measurements shown in Figure~\ref{fig:curv}-c to ensure that they are extracted from equilibrated domains. We find that $\Cpl$ depends linearly on the GM1 fraction as
\begin{equation}
\Cpl\simeq0.5 \, \phi_{\rm GM1}~{\rm nm}^{-1}
\label{eq:C1}
\end{equation}
This linear law is in agreement with the results of Ref.~\cite{Sreekumari2018} finding a comparable slope in their MD simulations, even though they have used another type of coarse-grained model. 

Below, we want to model non-planar systems (\textit{i.e.} vesicles) with our Mesoscale model. We denote by $C_0$ the spontaneous curvature of the Ld phase. In the present case, $C_0=2/R$ is set by the vesicle average radius $R$ in equilibrium, to accommodate the area difference between both leaflets~\citep{Hossein2020}. We recall that we also denote by $\Cvesi$ the spontaneous curvature of the Lo+GM1 phase on a vesicle. On a vesicle, we can also assume that before insertion of GM1, the Lo phase has the same curvature $C_0=2/R$, for the same reason as for the Ld phase, so that the GM1 curvature comes {\em in addition} to $C_0$ as examined in detail in Ref.~\cite{Hossein2020}. Thus $\Cvesi = C_0+\Cpl$ where $\Cpl$ is given in Equation~\eqref{eq:C1}.  A thorougher discussion about this relation is provided in the ESI$^\dag$.

\subsubsection{\textit{Determination of the Ising parameter through the measurement of the line tension}}
\label{sec:line_tens}

\begin{figure}[t]
\centering
\includegraphics[width=9cm]{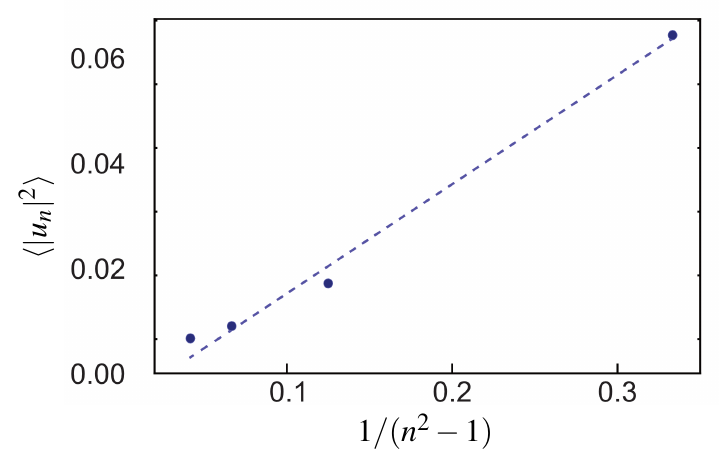} 
\caption{Power spectrum of the Lo-Ld boundary in a DPPC-DIPC-Chol mixture without GM1 from which the line tension $\lambda$ is extracted through a fit with Eq.~(\ref{eq:lambda}) (dashed line).}
\label{fig:lambda}
\end{figure}

As explained in the Method Section, one can relate the line tension $\lambda$ of the boundary between the Lo and Ld phase and the Ising parameter $J_I$ of the Mesoscale model. This relation ensues from the critical behavior of the mixture near its phase transition. Close enough to the critical value $J_{I,c}$, we have (\eq{eq:link2} ):
\beq
\lambda=\frac{4 \sqrt{3}}{a}(J_I-J_{I,c}).
\label{eq:link0}
\eeq
When one goes away from $J_{I,c}$, non-linear subdominant terms come into play~\ci{Feigenson2014}. We assume that this approximation provides reasonable estimates of $\lambda$ in the present case. Analytical~\citep{Baxter,Shneidman2001} or numerical solutions can also be used far from the critical point.

To measure the line tension at the boundary of Lo and Ld phases in a mixture devoid of GM1, we measure the fluctuations of the Lo domain boundary. We use the same method as the one applied on fluorescence microscopy images in \ci{Esposito2007} or in Mesoscale simulations~\ci{Ehrig2011}. 
We use the polar discretization (detailed in the Method section) in order to identify the location $r(\theta)$ of the Lo+GM1 domain boundary, the origin of coordinates being the domain barycenter. We then apply the 
DPPC fraction threshold of 0.6 (Figure~\ref{fig:compo_dist}-c) to delineate the boundary betwen Lo and Ld domains and calculate the Fourier coefficients $u_n$ of $u(\theta)$, defined by $r(\theta)=R_0[1+u(\theta)]$, using a Fast Fourier Transform. Finally, we use Eq.~(\ref{eq:lambda}) to relate the power spectrum of the boundary fluctuations to the line tension of the domain boundary $\langle|u_n|^2\rangle$ against $1/(n^2-1)$. We fit  only the first long wavelength modes to avoid discretization effects as shown in Figure~\ref{fig:lambda}. The fitted line tension is $\lambda \simeq 3$~pN for a DPPC-DIPC-Chol (30:58:12) mixture devoid of GM1 at $T=310$~K, of the same pN order of magnitude as experimental values~\ci{review}. 

We also measured how $\lambda$ depends on the GM1 concentrations, as shown in Table~\ref{tab:lambda}. In agreement with the CG simulations of Ref.~\cite{Liu2019}, $\lambda$ decreases when the GM1 concentration increases above 5\%, making the phase boundary more fuzzy. 
\begin{table}
\begin{center}
\begin{tabular}{lcccccc}
\hline
\% GM1            & 0 & 2.5 & 5 & 7.5 & 10 & 15 \\ \hline
$\lambda$ (pN)  & 3.0 & 3.0 & 5.0 & 2.4 & 1.2 & 0.8 \\ \hline
\end{tabular}
\end{center}
\caption{Measured line tensions $\lambda$ in function of the GM1 concentration in the upper leaflet.
\label{tab:lambda}}
\end{table}

\subsubsection{Determination of vertex-composition flips timescales}

Here and in the following section, we focus on dynamic questions that do not concern equilibrium properties of phase separation. Indeed convergence to equilibrium only relies on the detailed balance condition~\citep{Barkema}, and not on the exact values and physical relevance of the Monte Carlo transition rates in the simulation framework. The reader interested only by equilibrium issues can skip these two sections.

In the Mesoscale model, the rate at which vertex-composition flips (Kawasaki dynamics, see Methods) are attempted must be set by a relevant timescale $\delta t$ at the lengthscale $a$, the average lattice spacing. This is related to the diffusion coefficient $D$ of the minority species in the bulk of the majority phase at this scale, through the relation $a^2 = 4 D \delta t$. However, the effective value of $D$ also depends on the mesoscopic lengthscale $a$, as we discuss it now. Note that $D$ cannot simply be set as the diffusion coefficient of molecules in the bilayer because a membrane patch at the mesoscale can be a collection of dozens of lipids, or even more. Its diffusive properties are the fruit of the collective behavior of its interacting elementary constituents, themselves in interaction with the surrounding fluid. 

Domain boundary fluctuations are known to be a relevant probe of diffusion in phase-\-separated membranes~\cia{Esposito2007,Camley2010A,Camley2010B}. A notorious difference can already be noticed between \textcolor{black}{our} mesoscopic modeling on the one hand and MD simulations or experiments on the other hand. While hydrodynamic interactions are absent in the former by construction of \textcolor{black}{our} MC simulations, they are intrinsically present in the latter where the solvent is explicitly simulated
\textcolor{black}{(note however that accounting for hydrodynamic interactions in Monte Carlo simulations in general~\citep{Kikuchi1992} and in membrane modeling in particular is feasible in principle~\citep{Lin2004,Reister2007}, at least as far as homogeneous membranes are concerned)}. This implies different scaling laws of the timescales in function of the lengthscales. 

Refs.~\cite{Camley2010A} and \cite{Camley2010B} thoroughly address the role of hydrodynamic interactions in this context, in the frame of the Saffman-Delbruck theory. One can define a typical length, the Saffman-Delbruck length $L_{\rm SD}=h\eta_m/(2\eta_f)$, where $\eta_m$ and $\eta_f$ are the viscosities of the membrane and the fluid, respectively, and $h$ is the membrane thickness. 
$L_{\rm SD}$ is on the order of 100~nm to 10~$\mu$m in experimental systems. On lengthscales below $L_{\rm SD}$, 2D hydrodynamics inside the membrane dominates, while above it, 3D hydrodynamics in the surrounding solvent does. Considering a boundary fluctuation wavelength $\Lambda$, it follows that its relaxation time scales as $\tau(\Lambda) \propto \Lambda^2$ if $\Lambda \gg L_{\rm SD}$ and  $\tau(\Lambda)  \propto \Lambda$ if $\Lambda \ll L_{\rm SD}$. In the ESI$^\dag$, we properly define relaxation times and we discuss these relations. Below, we focus on (nanometric) values of $\Lambda \ll L_{\rm SD}$ at the molecular scale, so that 
\begin{equation}
\tau_{\rm MD}(\Lambda) = \frac{2 h \eta_m }{\pi\lambda}\Lambda
\end{equation}
where $\lambda$ still denotes the line tension.

Our choice to set the value of $D$ in Mesoscale simulations consists in identifying the molecular and mesoscopic time scales at the smallest wavelength accessible in the Mesoscale model, namely $\Lambda=2a$~\footnote{Indeed, the circumference of a circular domain of radius $R_0$ is discretized into $N_R \simeq 2\pi R/a$ elementary edges of the sphere tesselation. The discrete Fourier transform of the boundary fluctuations is written with Fourier coefficients $u_n$ associated with wavelengths $2\pi R/n$ (see Methods), $n$ running from $-N_R/2+1$ to  $N_R/2$. Hence the smallest accessible wavelength is $2\pi R/(N_R/2)=2a$.}. This means that we assume here that hydrodynamic interactions are at play up to the lengthscale $a$ only. Since hydrodynamics interactions are known to enhance dynamics~\citep{Manghi2006} it implies that the larger $a$, the faster the dynamics at large scales, thus the larger effective $D$, as we quantify it now. 

In the Mesoscale simulations where no hydrodynamic interactions are at play and where the order parameter is conserved, it can be demonstrated~\citep{Destain2022} that the relaxation time at wavelength $\Lambda$ is related to the diffusion coefficient $D$ through 
\begin{equation}
\tau_{\rm Meso}(\Lambda) = A \frac{\lambda }{D J_I} \Lambda^3 
\end{equation}
where $A$ is a dimensionless constant that was determined numerically on the triangular lattice, $A \simeq 0.1$. The identity $\tau_{\rm MD}(2a) = \tau_{\rm Meso}(2a)$ finally sets the value of the diffusion coefficient $D$, and therefore the timescale $\delta t=a^2/(4D)$ under interest. Here we need the value of $J_I$, which also depends on $a$. It can be estimated through the approximation in Eq.~\eqref{eq:link0} where $J_{I,c} = \frac{ln 3}{4}  k_{\rm B}T$. We eventually get
\begin{equation}
D \simeq 2 \pi A\frac{\lambda^2 a^2}{h \eta_m} \frac4{ k_{\rm B}T\ln 3+a \lambda / \sqrt{3}}
\label{eq:D}
\end{equation}
As expected, this diffusion coefficient $D$ in the membrane plane depends on $a$. It also depends on $\lambda$ that can be measured as explained above. We emphasize that this relation simply reflects our choice to get the timescales at lengthscale $2a$ to coincide. In this expression, the line tension $\lambda$ expresses the role of the interactions between the molecular constituents. 

Given that dynamics are known to be significantly enhanced in MD simulations using the MARTINI force field~\cia{GROMACS,denOtter2007,Vogele2019} (see also the ESI$^\dag$), we prefer to use experimental values of $\eta_m$ rather than numerically measured ones. In the present case, we estimate in the SI that $h \eta_m \simeq 4 \times 10^{-10}$~Pa.m.s. For instance, for a vesicle of radius $R=30$~nm as considered below, the average tessellation edge length is $a=2.25$~nm for $N=2562$ vertices. With the measured value $\lambda\approx 0.8$~pN at 15\% of GM1 in the upper leaflet, and $T=310$~K, we get the numerical values $D \approx 3.6$~$\mu$m$^2$/s and finally $\delta t \approx 0.35$~$\mu$s. The value of $D$ is lower than lipid diffusion coefficients measured in real model membranes, e.g. vesicles~\citep{Lindblom2009}, being however on the same order of magnitude because with this value of $R$, patches are small, each vertex representing about 6 lipid molecules in each leaflet.

\subsubsection{Determination of transverse deformation modes timescales}
Now we address the question of the transverse fluctuation timescale in the Mesoscale model. A well established theory~\cia{Seifert1993,Seifert1997} provides the membrane friction constant per membrane unit area 
at wavevector $q=2\pi/\Lambda$, 
\begin{equation}
\zeta_\perp = 4 \eta_{\rm f} q = \frac{8 \pi \eta_{\rm f}}{\Lambda}
\end{equation}
We are again interested in vertex dynamics at the shortest wavelength $\Lambda=2a$. Owing to Einstein's relation, the radial diffusion coefficient of an elementary membrane patch of area $A_0 = 4 \pi R^2/N \sim a^2$ is thus $D_\perp = k_{\rm B}T /(\zeta_\perp A_0)$. In our Mesoscale simulations, radial MC moves are attempted with a spatial step $\delta r = \rho R$ (with $\rho=0.007$ in most of our runs). Thus the physical timescale associated with (one-dimensional) radial MC moves is 
\begin{equation}
\delta t_\perp = \frac{\delta r^2}{2 D_\perp} =
\frac{2 \pi \eta_{\rm f} \delta r^2 A_0}{k_{\rm B}T a}
\end{equation}
If $R=30$~nm, we obtain $\delta t_\perp=90$~ps. Compared to the time-scale $\delta t$ associated with vertex-composition flips above, it is much shorter. It means that if one were interested in studying realistic vesicle dynamics through kinetic Monte-Carlo simulations~\citep{Barkema}, one would need to execute a large number of vertex radial moves in-between two vertex-composition flips, on the order of $\delta t / \delta t_\perp \approx 4000$. This value depends on the value of  the vesicle radius $R$ through several parameters.

\subsection{Mesoscale simulation}

Relying on the numerical results at the coarse-grained scale and after the extraction of the key parameters, we perform the Mesoscale Monte Carlo simulation of a tessellated vesicle made of two phases, one corresponding to the Ld phase, the other one, with a higher spontaneous curvature, accounting for the Lo phase curved by GM1 insertions. We simulate a vesicle with $N=2562$ vertices. Since our objective is to come back later to the CG scale through backmapping, we shall focus here on a small unilamellar vesicle (SUV). Details of the Mesoscale numerical scheme are given in Refs.~\cite{GG2} and~\cite{Cornet2020}, and summarized in the Methods section. 

\begin{figure}[t]
\centering
\includegraphics[width=9cm]{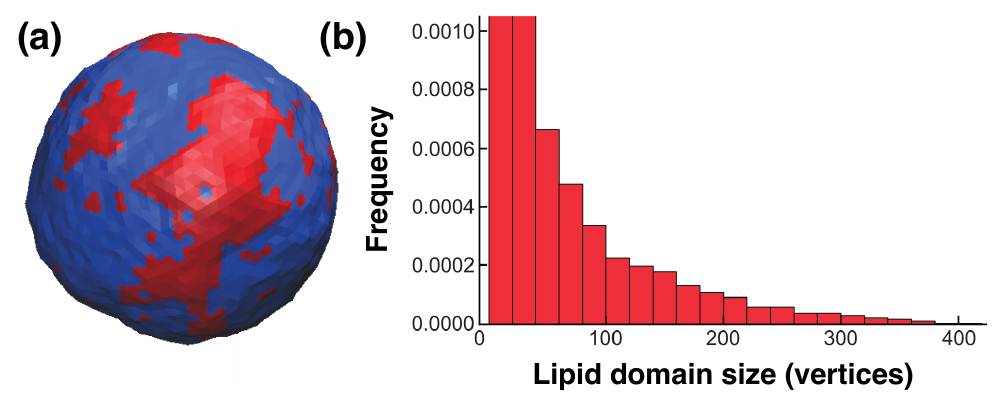}
\caption{Lipid nano domain sizes for the Mesoscale model. (a) snapshot of a mesopatterned SUV of radius 30~$\mu$m after equilibration through the Mesoscale model. The area fraction $\phi$ of the Lo+GM1 curving phase (in red) is equal to 20\%, that is to say there are 512 red vertices out of 2562 (each vertex is represented here by 6 small triangles defining its Voronoi cell. For one given vertex, the 6 triangles are all either red or blue). Other parameters are given in the text. {(b)} nanodomain size distribution. Nanodomain sizes are given in number of vertices in a given (Lo+GM1)-phase nanodomain. The two first bars have been truncated for clarity.}
\label{fig:backmap1}
\end{figure}

As notably discussed by Ref.~\cite{Cornet2020}, coarse-graining up to the mesoscale is a quantitative process which consists in pre-averaging the degrees of freedom at length-scales smaller than the lattice spacing $a$, the physical foundations of which are prescribed by the renormalization group theory~\citep{Chaikin,Feigenson2014}. Consequently, one expects the Mesoscale model to describe correctly the fluctuations of the real system at large scales, with the advantage that it will reach equilibrium much faster in terms of computational time, provided, as described just above, that microscopic details have been correctly integrated out in the Mesoscale  parameters. Note that in the same spirit, force fields used in all-atom or CG models have already integrated out quantum mechanics degrees of freedom, and are consequently already approximate. 

As explained above, we use the following parameters values to simulate a vesicle of real radius $R=30$~nm with the same Ld and Lo+GM1 phases as in the bilayer with 15\% of GM1 in the upper leaflet studied above:
\begin{enumerate}
\item The bending moduli are set to $\kLo=13k_{\rm B}T$ and $\kLd=25 k_{\rm B}T$ after calculation of the membrane fluctuation spectra in the CG model.
\item The spontaneous curvature of the Ld phase is equal to $C_0=2/R$, the one of the average sphere. The spontaneous curvature of the Lo+GM1 phase is fixed to $\Cvesi=C_0+\Cpl=4.43/R$, from measurements of $\Cpl$ in CG simulations at 15\% of GM1, as explained above. 
\item The Ising parameter is set to $J_I=0.336 k_{\rm B}T$ to match the measured line tension $\lambda$, through Eq.~\eqref{eq:link0}. This value is above the critical one $J_{I,c} = (\ln 3/4) k_{\rm B}T \simeq 0.275 k_{\rm B}T$ on a triangular lattice.
\item The surface tension is arbitrary since it is an extrinsic parameter imposed by experimental conditions, and not a property of the membrane. In the simulations, its dimensionless value $\tilde \sigma =  \sigma R^2 /(k_{\rm B} T)$ was set at 1100, which corresponds to $\sigma = 5.3 \times 10^{-3}$~J/m$^2$ for a radius  $R=30$~nm. This value ensures that the quasi-spherical vesicle regime, where our Mesoscale model is fully valid~\citep{GG2}. 
\item The area fraction $\phi$ of the Lo+GM1 phase is chosen equal to 20\%, so that Lo domains are well-defined. Increasing this value would lead to more complex interdigitated, labyrinthine patterning~\citep{Cornet2020}. 
\end{enumerate} 
The system was run for a long time ($10^{10}$ MC steps) to ascertain that thermodynamic equilibrium has been reached~\citep{Cornet2020}. A snapshot by the end of the simulation is displayed in Figure~\ref{fig:backmap1}. 

With the parameters inferred from the CG model, the SUV displays nanodomains in equilibrium. As observed in previous work~\citep{Cornet2020}, the domains form rapidly after starting the simulation from a configuration where Ld and Lo vertices were randomly distributed. Then the domains fluctuate in size and shape, permanently exchanging Lo ``monomers'' with the surrounding phase. They can also coalesce or split. It must be emphasized that as compared to the CG simulations above where a single nanodomain, necessarily smaller than the simulated box size, was formed in few $\mu$s on the quasi-flat membrane due to phase separation, the full phase separation displaying a single, large Lo domain, is avoided on the curved SUV, by introducing a surface tension, $\sigma$. Hence, the nanodomain size is controlled by the typical length scale $\lambda/\sigma$, since larger curved domain have a too large surface free-energy~\cite{review}. The nanodomains are not generically larger (in real size) than in the CG simulations, but they never coalesce in a single, large Lo region.

To quantify this vesicle meso-patterning, we have computed the nanodomain size-distri\-bution, as illustrated in Figure~\ref{fig:backmap1}. As compared to distributions generally observed on large GUVs  below the critical temperature (see ESI or Ref.~\cite{Cornet2020} for examples), the distribution is not bimodal but monotonously decreasing. The main reason for this difference is that on a SUV, the ratio between the Lo domain curvature $\Cvesi$ and the sphere one $C_0$ is moderate as compared to a GUV.
The Mesoscale model thus predicts that nanodomains on SUVs can dynamically adopt a wide range of sizes, from very small to very large (with respect to the number of available Lo vertices) ones, that however never reach the maximum allowed size of 512.

\begin{figure}[t]
\centering
\includegraphics[width=9cm]{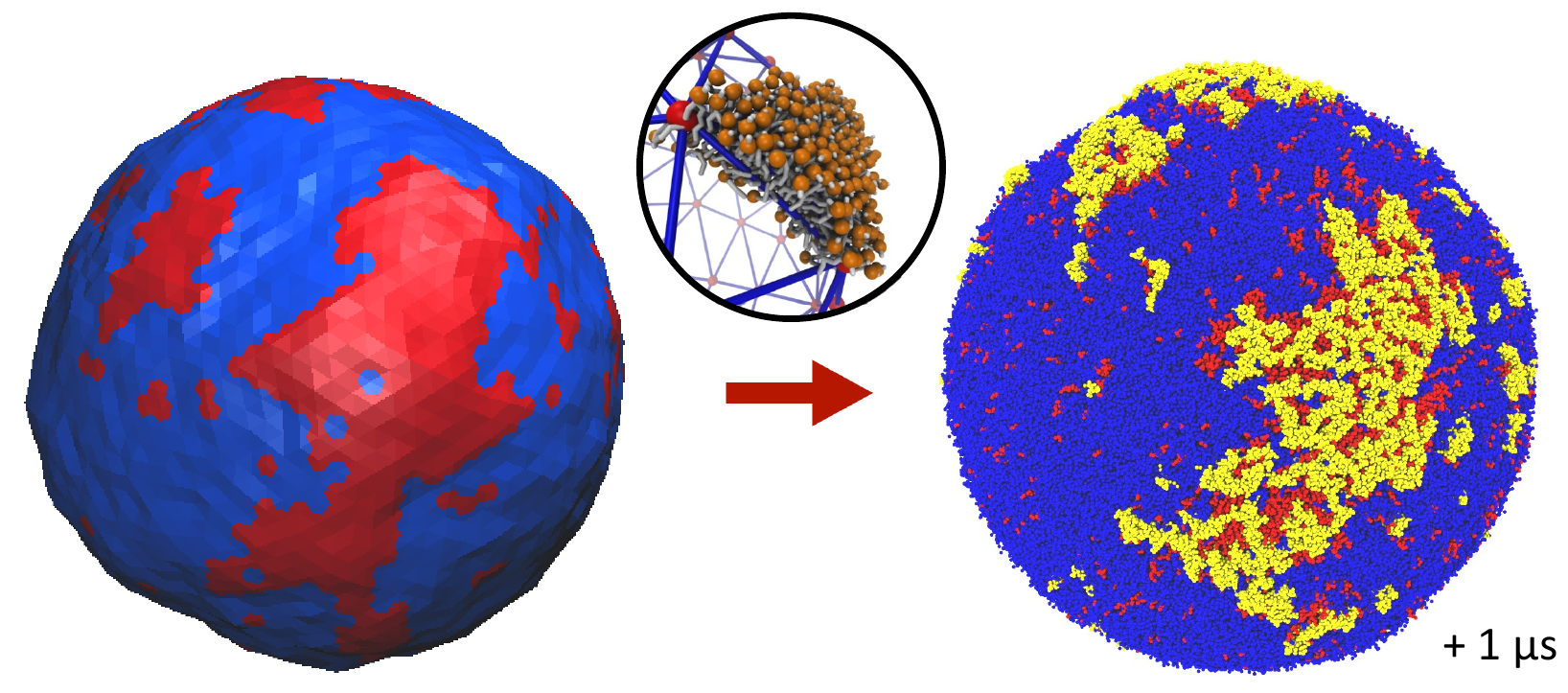} 
\caption{Backmapping from the Mesoscale model (left, equilibrated vesicle) to the CG one (right, after 1~$\mu$s of CG simulation). The central medaillon illustrates how each triangle of the tessellated vesicle is filled with a lipid bilayer patch. 
\label{fig:backmap2}}
\end{figure}

\subsection{Backmapping from the Mesoscale model to the CG one}

\begin{figure}[t]
\centering
\includegraphics[width=9cm]{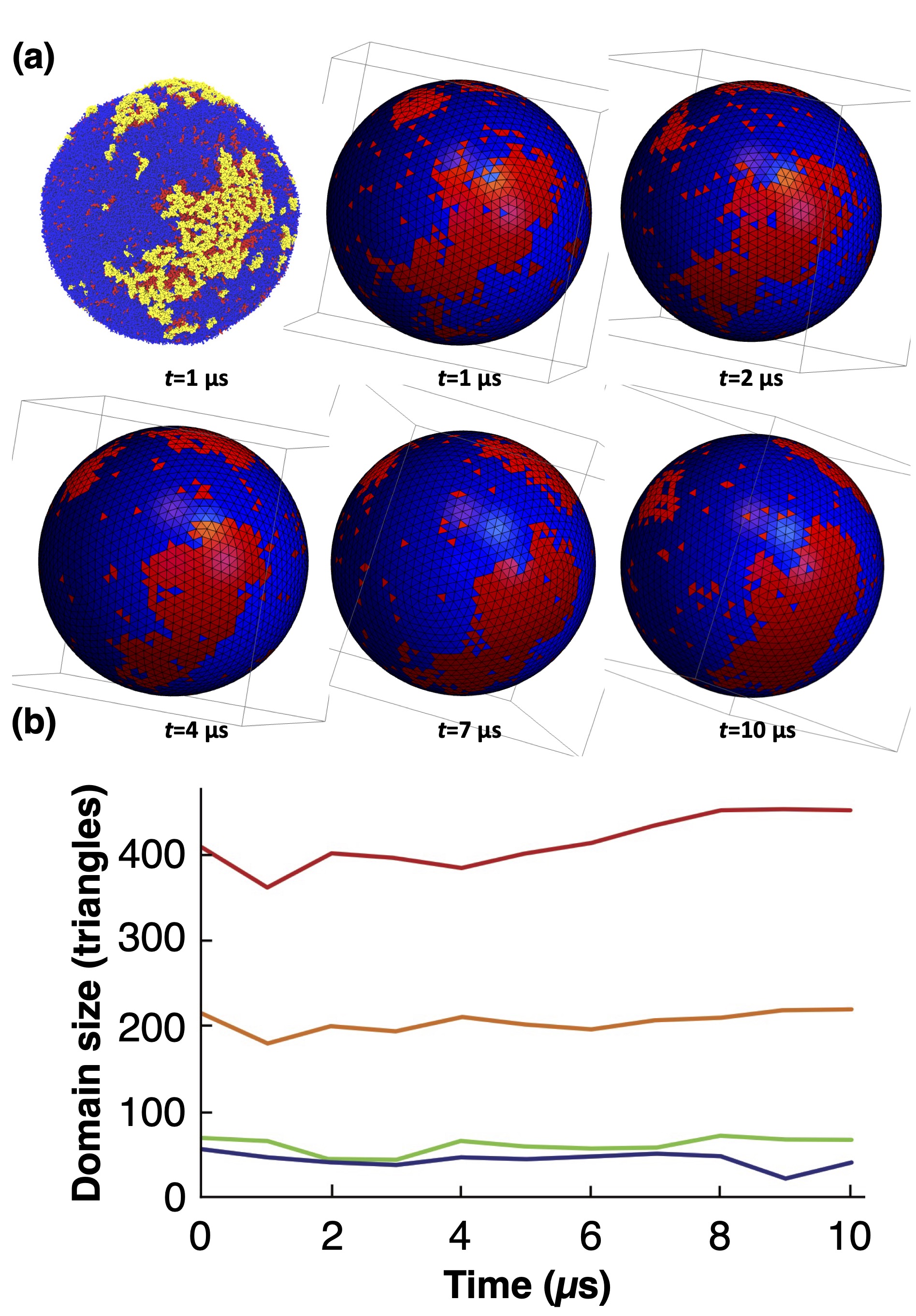} \\
\caption{Lipid Lo domains evolution in the CG-MD simulation. (a) CG-simulated vesicles projected onto the sphere and discretized again through the majority rule, at different simulation times. The first panel shows the CG vesicle at $t=1~\mu$s, for comparison. The other frames have been rotated  so that one can track the same lipid domains even though they diffuse slowly on the vesicle. (b) Domain size evolution with time (4 largest domains only). Note that the vesicle has twice as many triangles as vertices, so the largest domain, visible on the successive snapshots, contains about 200 vertices out of $N=2562$. }
\label{fig:backmap3}
\end{figure}

Backmapping from mesomodel to CG system consists in decorating the elementary triangles of the Mesoscale tessellation while respecting both the bilayer geometry and its local composition~\cite{Pezeshkian2020} (see Methods). The triangles are decorated with the same Ld and Lo+GM1 phases as in the bilayer with 15\% of GM1 in the upper leaflet studied above \textcolor{black}{with the help of CG simulations}. The interior and exterior of the $R=30$~nm-radius vesicle are filled with water (and ions) in order to constraint the vesicle interior volume as in the Mesoscale simulation. Backmapping is exemplified in Figure~\ref{fig:backmap2}. We shall now verify that after backmapping from the Mesoscale model to the CG one, CG simulations run from the so-obtained simulation output are stable through time, which will support the fact that the Mesoscale parameters were correctly estimated. Whereas simulations at the mesoscale enable one to equilibrate the slow long wave-length fluctuations, above the lattice spacing $a$, at the price of ignoring the short wavelengths, CG simulations after backmapping enable one to equilibrate in turn these short, fast wavelengths, and finally to obtain a CG system equilibrated at all wavelengths, whereas equilibrating long wavelengths with the help of the sole CG model would be prohibitive in terms of computational cost.

In order to assess the stability of the backmapped CG vesicle, notably in terms of its nano-patterning, we have tracked the largest nanodomains along a 10~$\mu$s-long CG simulation, as shown in Figure~\ref{fig:backmap3}-a. In order to identify unambiguously the Lo nanodomains, we have radially projected the lipid coordinates on a regularly tessellated sphere made of 5120 elementary triangles and attributed the Lo or Ld nature to each of them by using the same kind of majority rule as previously, illustrated in Figure~\ref{fig:compo_dist}. We have measured that the total number of Lo triangles remains remarkably stable through time, very close to the initial 20\% fraction before backmapping (see Fig. S6 in the ESI$^\dag$). Furthermore, nanodomains are identified as the connected components of the Lo phase. As expected, the size of the largest nanodomains fluctuates, as they regularly lose or gain lipids of the Lo phase, however their size does not evolve much with time (see Figure~\ref{fig:backmap3}-b). This suggests that the nanopatterning ensuing from the Mesoscale simulation remains qualitatively stable with time and validates our whole multiscale scheme.

\section{Discussion}

This work proposes a complete strategy to extract valuable parameters from coarse-grained simulations to design a Mesoscale model of biphasic vesicles. Going from the molecular scale to mesoscopic ones can present intrinsic difficulties, that we have fully addressed here, appealing to important physics concepts such as the renormalization group theory. This concerns in particular the scale-dependent relationship between the line tension at the interface between both phases, which is measured in CG simulations through the fluctuation spectrum of the interface between both phases, and controlled by the Ising parameter $J_I$ in the Mesoscale model. Once equilibrated, the mesoscopic system is back-mapped to the molecular scale through a well-defined procedure. This multiscale simulation scheme eventually allows one to obtain large systems equilibrated at all length scales, which could not be achieved at the sole CG scale owing to the prohibitive simulation time it would require. In particular, modeling at the Mesoscale enables equilibration of long length scales  before equilibration of short ones after back-mapping to the molecular scale. Here we have tackled a SUV of modest size (diameter of 60~nm) as a proof of concept, but there is no obstacle in principle to tackle larger systems. 

In the present context, and in relation with the experiments of Ref.~\cite{Shimobayashi2016}, our CG simulations gave us new insights into the effect of insertion of the glycolipid GM1 into the Lo phase. Since GM1 is only inserted in one leaflet, the ensuing asymmetry imposes a local spontaneous curvature to the membrane, proportional to the concentration of GM1 in the leaflet. In addition, GM1 makes the Ld/Lo interface more fuzzy, which is the visual manifestation of the decrease of the interface line tension $\lambda$. From a dynamical point of view, it also slows down the convergence to equilibrium. The two latter phenomena are intimately related since boundary fluctuation time-scales are inversely proportional to the line tension~\citep{Destain2022}. Increasing the spontaneous curvature of the minority phase while decreasing the line tension are both favorable to the stabilization of mesophases, where phase separation is incomplete in equilibrium, as was indeed observed in Mesoscale simulations. Note that the spontaneous curvature values imposed by GM1 are comparable to the spontaneous curvatures locally imposed by the insertion of some transmembrane proteins, measured by experimental or numerical techniques to be on the order of few 0.01~nm$^{-1}$ \citep{Aimon2014,Rosholm2017,Kluge2022,Noguchi2023}. Thus the meso-patterning effect observed at relatively large concentrations of GM1 is also expected to occur on cell membranes where curving proteins are abundant~\citep{review}.

Here we have used the Martini 2.2 forcefield. Recently, a new version of this forcefield was released increasing the number of bead types and sizes \citep{Souza2021} refining diverse features especially related to protein. This resulted in a recent reparametrization of carbohydrates \citep{Grunewald2022}  \textcolor{black}{but CG model of GM1 lipid has not been yet released}. It will be interesting, in the coming years, to extend our CG modelling to see how this new version affects the quantitative results. In the meantime, the Martini 2.2 version had given already reasonable results \citep{Liu2019,Dasgupta2018} on GM lipids and curvature to validate our proof of concept.

To perform our Monte Carlo Mesoscale simulations, we used a model that some of us developed and characterized in recent works~\citep{GG2,Cornet2020}. A more common model used in numerical Statistical Mechanics of membranes is \textcolor{black}{the Dynamically Triangulated Surface/Membrane (DTS/DTM) model \citep{Siggel2022,PezeshkianIpsen2023}}. Usually designed in the early 1990's to study the crumpling phase transition of homogeneous fluid membranes~\citep{Ho1990,Kroll1992}, it has been extended and enriched in order to model membranes or vesicles closer to cell membranes, in particular by endowing it with multiphasic composition~\citep{Hu2011,Pezeshkian2019}. In our model, the vertices of the tessellation experience only radial moves and there are no edge-flips, which are an important characteristic of the DTS model. 

These features do not modify in depth the overall properties of the model, however using our own model presents at least two advantages in the present context. First, as far as the Ising model is concerned, it dwells on a triangular lattice (except the 12 vertices that have only 5 neighbors, necessary to close the membrane, owing to Euler's polyhedron formula), whereas in the DTS case, it would dwell on a disordered lattice, with vertices that can have in principle any number of neighbors larger than 3. The rigorous connection between the line tension and the Ising parameter $J_I$ relies on the knowledge of the critical parameter $J_{I,c}$ and of the prefactor $4 \sqrt{3}$ in Equation \eqref{eq:link0}, 
whereas they are not exactly known on a disordered lattice. The same problem arises when it comes to diffusion and dynamical issues, as in Equation~\eqref{eq:D}. Second, in the DTS model, edge lengths are constrained by an upper and a lower bound, in order to ensure membrane self-avoidance and to prevent some triangles to become very large to the detriment of the smallest ones. Because of entropic effects due to these constraints in the configuration space, an additional entropic term modifies the input surface tension $\sigma$ in a way that is difficult to control because it depends in a non-trivial way on the total area of the vesicle. The quantification of this effect is out of the scope of this work and will be the object of a forthcoming publication. For these reasons, we did not opt for the DTS model in the present study where every mesoscale parameter must be well-controlled. 

The membrane surface tension is one of the key parameters of the Mesoscale model because the topology of the meso-pattern~\citep{komura2006} and even its stability depend on the surface tension value. A large enough value is required to destabilize the macrophase whereas a too large value would suppress membrane fluctuations and thus also lead to macrophase separation~\citep{Weitz2013,review,Cornet2020}. When backmapping to CG simulations, the surface tension can easily be monitored in planar geometry by applying anisotropic pressure on the simulation box~\citep{GROMACS}. However, controlling the surface tension is more challenging in vesicle geometry~\citep{Ollila2009}. Indeed, it depends on the quantity of water and ions encapsulated inside the vesicle, owing to the Laplace law, through the difference of pressures across the membrane. In this work, we have not quantified the surface tension of the vesicle after backmapping, but simply filled the vesicle with the maximum quantity of water (plus ions), anticipating that since the overall interior volume was the same as in the Mesoscale vesicle simulations, the surface tensions should be comparable. Since it is difficult to compute \textit{a priori} the value of the surface tension, the most obvious way to control it is to proceed by trial and error, removing water and ions if the tension it too high and adding ones if it is to weak, for example by pushing the existing solvent away from the center of the vesicle through application of a radial force. This issue will be addressed in a near future. 

In the era of exascale computing \citep{wieczor2023,Chang2023}, it is tempting to design molecular systems to study from organelles \citep{Pezeshkian2020,Gupta2022} to viruses \citep{Casalino2023,Gupta2022} or even an entire cell \citep{Vermaas2021,Lutheyschulten2022,Stevens2023}. Yet, the computing power necessary to simulate such large systems is still very huge and prevents one from reaching simulated timescales useful to decipher even simple phenomena such as diffusion of large molecules \citep{Gupta2022} without even mentioning relevant biological time scales of milliseconds or longer. Our approach may help tackling this issue by equilibrating a simplified triangulated version of such huges systems using Monte Carlo simulations to then only refine the equilibrated system on nano- to microseconds time scales.

\section*{Acknowledgments}
 M.C. is supported by the CNRS-MITI grant “Modélisation du vivant” 2020 and by the ITMO Cancer of Aviesan “Mathematiques et Informatique”. This work was granted access to the HPC resources of CALMIP supercomputing center (under the allocation 2021-P21020) and TGCC Joliot-Curie supercomputer (under the GENCI allocation A0110712941). \textcolor{black}{W.P. acknowledges funding from the Novo Nordisk Foundation (grant No. NNF18SA0035142) and Independent Research Fund Denmark (grant No. 10.46540/2064-00032B).}

\section*{Author Contributions}
J.C. performed the simulations and analyzed the results. N.C. contributed to the analyses and M.P.-M. contributed to the CG simulations.  M.C., W.P. and S.J.M. contributed to the simulations and result analyses. M.M., M.C. and N.D. designed and supervised the research. J.C., N.D., M.C. and M.M.  wrote the manuscript.

\section*{Conflicts of interest}
There are no conflicts to declare.






\scriptsize{
\bibliography{rsc} 
\bibliographystyle{rsc} } 


\setcounter{equation}{0}
\setcounter{figure}{0}
\setcounter{table}{0}
\setcounter{section}{0}
\renewcommand{\figurename}{Fig.~SI.}
\renewcommand{\tablename}{Table~SI.}

\newpage

\normalsize 

\begin{center}
{\Large \textbf{Supplementary Information to the paper
\\``There and back again: bridging meso- and nanoscales to understand lipid vesicle patterning''}}\\ 

\medskip 

Julie Cornet, Nelly Coulonges, Weria Pezeshkian, Maël Penissat-Mahaut, Siewert-Jan Marrink, Nicolas Destainville, Matthieu Chavent, and Manoel Manghi

\end{center}

\section{Cumulative effects of spontaneous curvature induced by insertions and by area difference}

Spontaneous curvature can arise from either local molecular insertions with asymmetric shapes, such as lipids or proteins with global conical shapes, or from the  difference of area between the two leaflets of the bilayer. This difference of areas comes from the history of the bilayer, the way its was created and equilibrated (or not). In this case, it is a non-local effect, at the origin of the Area Difference Energy (ADE) model~[\ref{Seifert1997},\ref{Mouritsen2005}]. In Ref.~[\ref{Hossein2020}], the cumulative effects of both curvatures is investigated. It is shown (see their Eq. (18)) that the resulting spontaneous curvature can be written as the weighted mean 
\begin{equation}
C^* = \frac{\kappa K_{\rm B} + \kappa_{nl} K_s}{\kappa + \kappa_{nl}} 
\end{equation}
where $K_{\rm B}$ and $K_s$ are curvatures associated with asymmetric insertions and area difference, respectively, and $\kappa$ and $\kappa_{nl}$ are two bending moduli. It is argued in this work that $\kappa_{nl}  \gg \kappa$ (a ratio of 6 is proposed) so that 
\begin{equation}
C^* \simeq \frac{\kappa}{\kappa+ \kappa_{nl}}  K_{\rm B} + K_s
\end{equation}
When both leaflets have the same area, as in our planar geometry CG simulations, one has $K_s=0$ so that the measured spontaneous curvature in presence of GM1 is  $\Cpl = \frac{\kappa}{\kappa+ \kappa_{nl}}  K_{\rm B}$. In vesicles where leaflet areas have had time to accommodate the spherical geometry, $K_s = 2/R$. We eventually get that when both effects are concomitantly at play, $C^*=\Cpl +2/R$. This $C^*$ corresponds to our $\Cvesi$ in the main text. In other words, the curved species is endowed with a differential curvature $\Cpl$ measured in planar geometry.

\section{Estimate of bending moduli $\kappa$}

\begin{figure*}[t]
\centering 
\includegraphics[height=4cm]{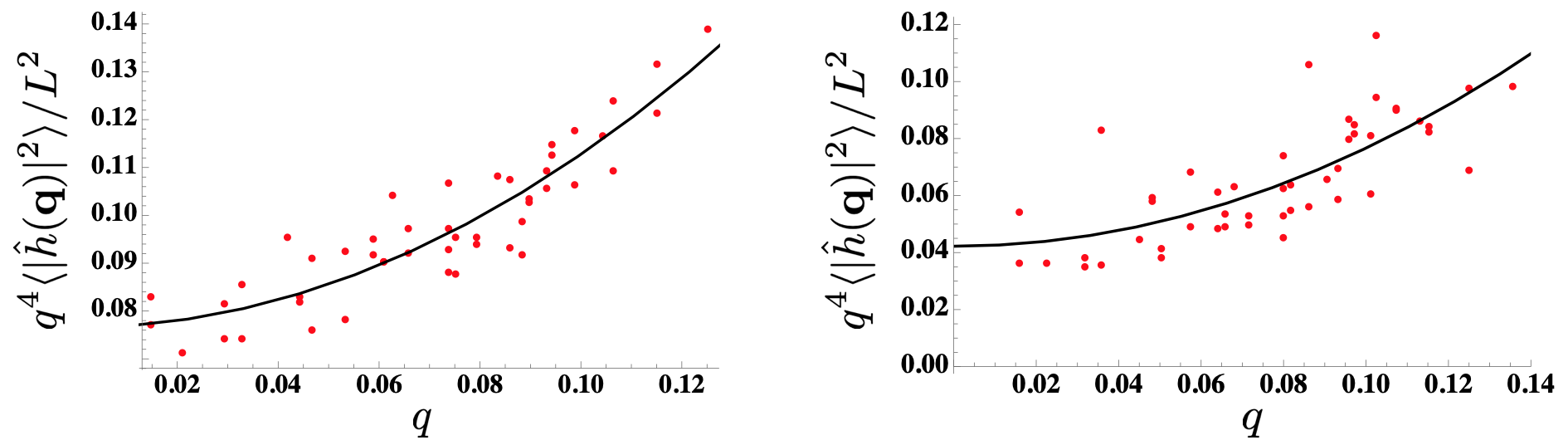} 
\caption{Spectral densities (dots) from CG 10~$\mu$s simulations of $L\approx 40$~nm square patches. Ld (left) and Lo+GM1 (right) lipid mixture bilayers were simulated and the spectra fitted with Eq.~(12) of the main text (black line), from which bending moduli are extracted. The wavevectors $q$ are expressed  in $\AA^{-1}$ and the vertical axis unit is $(k_{\rm B}T)^{-1}$.}
\label{fig:fluct_spectra}
\end{figure*}

We follow the strategy proposed by Ref.~[\ref{Fowler2016}] to extract the bending modulus $\kappa$ from a CG simulation of a tensionless bilayer made of a homogeneous lipid mixture. As made explicit in Eq.~(12) of the main text, the membrane thermal shape fluctuations can be related to membrane parameters through Helfrich's model, to which a term accounting for protrusions has been added. To measure the Fourier mode amplitudes from CG simulation, we used the PO4 beads of phospholipids to identify both leaflets and used the Mathematica interpolation function (at order 1) to transform their positions into a continuous height function representing the membrane mid-surface, taking periodic boundary conditions into account, before Fourier-transforming it. Fitting with Eq.~(12) of the main text, we got $\kappa_{\rm Ld}\simeq 13.1 \pm 0.3 k_{\rm B}T$ and $\kappa_{\rm Lo}\simeq 24\pm2 k_{\rm B}T$ (standard error are provided by the fitting procedure). We also found the value $\sigma_{\rm pr}\simeq 0.12 $~J/m$^2$ for the tension associated with protrusions, in agreement with expectations. In Figure~\ref{fig:fluct_spectra}, we observe that the Lo data are more noisy than the Ld ones, even though both CG simulations were of equal duration. This might be due to the presence of GM1 bulky heads perturbing the bilayer order. For this reason, we round the measured value to $\kLo = 25 k_{\rm B}T$ for the Lo phase. We also round the value to $\kLd = 13 k_{\rm B}T$ for the Ld phase. 

Now we compare these values to available experimental ones found in the literature. Indeed, many experimental techniques can also give access to the value of the bending modulus of a homogeneous phase, notably micropipette aspiration, measurements of vesicle shape fluctuations (flicker spectroscopy), NMR, electrodeformation, low angle diffuse X-ray scattering or neutron spin-echo. Note however that the so-obtained experimental values can appear to depend significantly on the experimental method used~[\ref{Doktorova2017}]. For a pure DPPC membrane explored by flicker spectroscopy just above the transition temperature, one measures a bending modulus on the order of $10k_{\rm B}T$~[\ref{Fernandez1994}]. The adjunction of 30\% cholesterol increases this value by a factor $\sim2$ at $44^\circ$C~[\ref{Oradd2009}], as measured by NMR. This is consistent with the value $\kLo \simeq 25k_{\rm B}T$ found with our CG simulations, assuming that GM1 do not affect too much the Lo phase rigidity, being a minority species. For a pure DLiPC (DIPC in MARTINI)  membrane, its bending rigidity has also been measured to be close to  $10k_{\rm B}T$ by micropipette aspiration at 18$^\circ$C~[\ref{Rawicz2000}], also comparable to the value that we measured. 

\begin{figure}[t]
\centering 
\includegraphics[height=4.5cm]{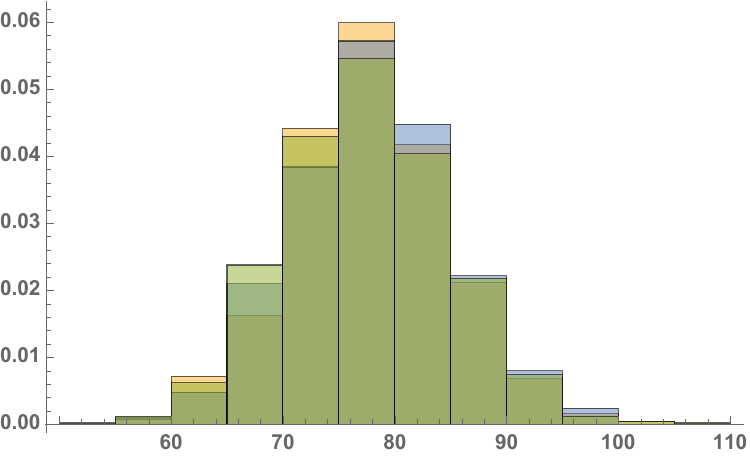} 
\caption{Influence of the rigidity $\kLd$ of the Ld phase on the nanodomain size distribution for $\kLd = 6$ (green bars), 10 (blue) and $20 k_{\rm B}T$ (yellow). Nanodomain sizes are given in number of vertices in a given Lo-phase nanodomain in the Mesoscale model. The distribution is bimodal, however the monomer and small multimer peaks are not shown. Other parameter values are $\phi=0.2$, $\kLo = 20 k_{\rm B}T$, $\Cvesi=5/R$, $J_I=1/3.4>J_{I,c}$, and $\tilde\sigma=300$.}
\label{fig:var:kappa}
\end{figure}

Finally, we checked that the exact choice of this latter value has no strong influence on the nanodomain statistics in the Mesoscale model because the Ld phase is assumed to have its spontaneous curvature set by the average vesicle radius and is only weakly curved. For example, in Fig.~\ref{fig:var:kappa}, we give the nanodomain size distributions for three different values of $\kLd$ and see little influence of this parameter value. 

\section{Local thickness measurements}

\begin{figure*}[t]
\centering 
\includegraphics[width=14cm]{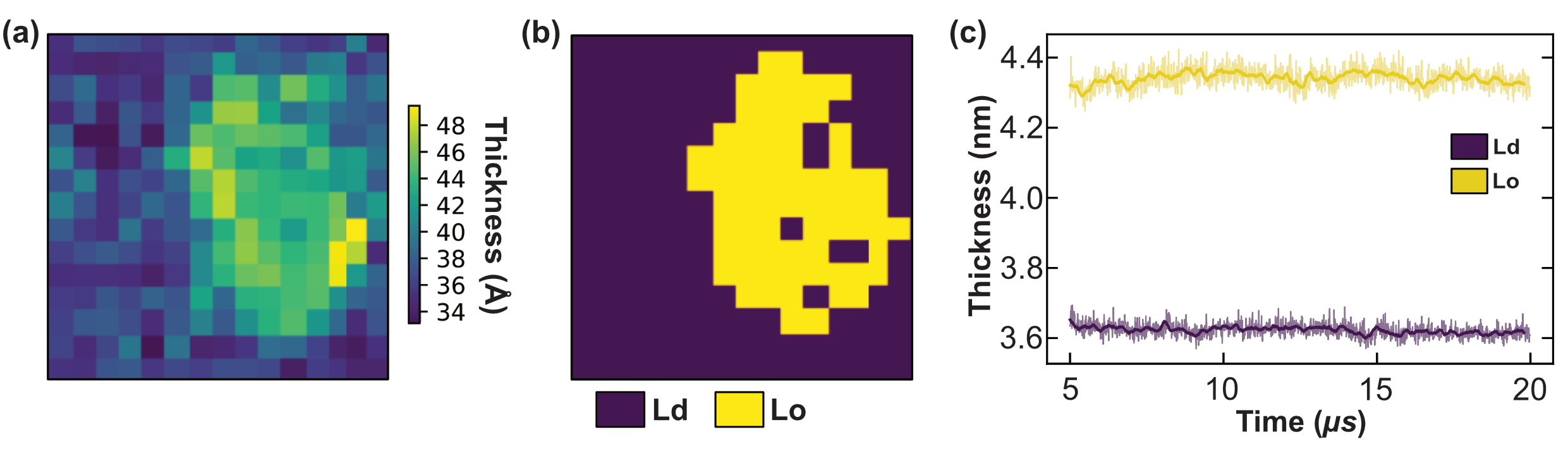} \\
\caption{{\bf (a)} Local thickness $h$ measured on the $15 \times 15$ mesh (in \AA), compared to the binarized membrane {\bf (b)}, for the same mixture as in Fig.~5 in the main text. {\bf (c)} Local thickness (in nm) evolution with time after equilibration, measured separately for Lo and Ld phases. DPPC-DIPC-chol (30:58:12) mixture with no GM1. Here and in the following figures, the thick lines are sliding averages of the thinner ones on a 0.2~$\mu$s sliding window.}
\label{fig:thick}
\end{figure*}

The local thickness can be calculated via the difference of position of the corresponding box centroids in the two leaflets, for the Ld and the Lo phase (Fig.~\ref{fig:thick}). We find an average membrane thickness for all GM1 concentration of approximately 4~nm. As expected, the Lo phase is thicker than the Ld one. This explains why the Lo phase is more rigid than the Ld one. Indeed, in rough approximation, a membrane can be seen as an homogeneous elastic plate, the bending modulus of which grows as the cube of its thickness $h$:
\begin{equation}
\kappa = \frac{E h^3}{12 (1 - \nu^2)}
\end{equation}
assuming the same Young modulus $E$ and Poisson ratio $\nu$~[\ref{Landau1986}]. Using the average values shown in Fig.~\ref{fig:thick} (bottom), one finds $\kLo/\kLd\simeq  (4.35/3.6)^3\simeq 1.8$ in full agreement with our measures of $\kLd$ and $\kLo$. 

\section{Correlations}

\begin{figure*}[t]
\centering
\includegraphics[width=14cm]{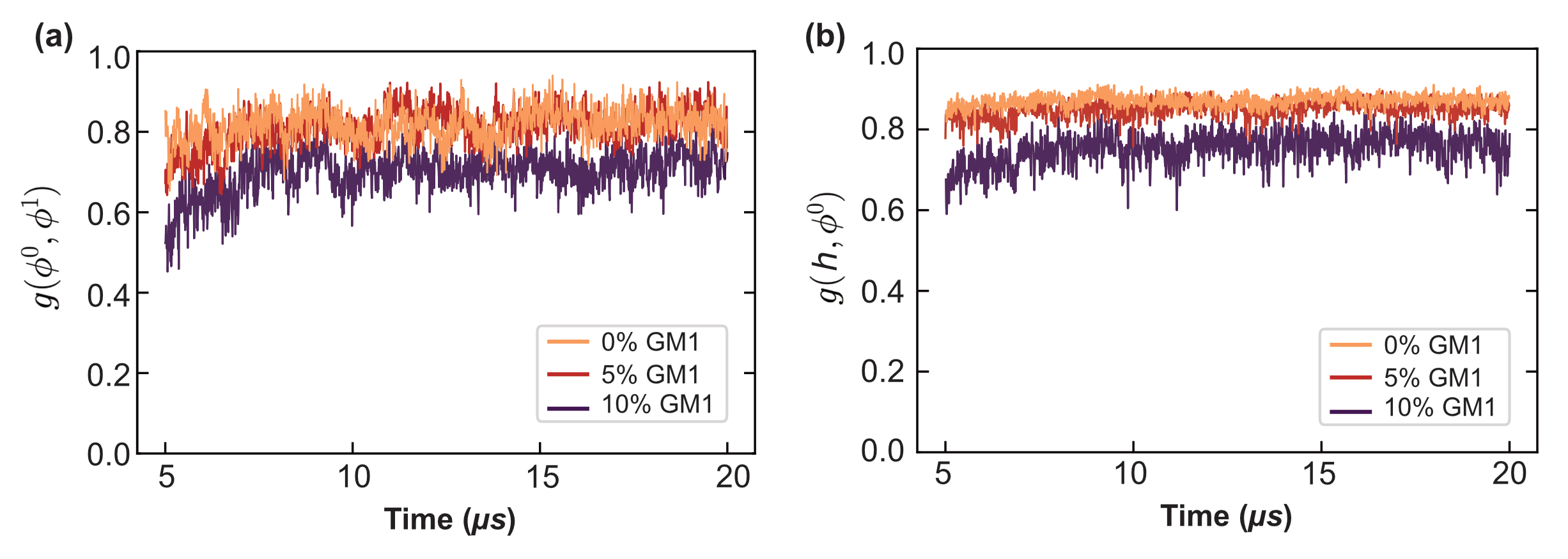} 
\caption{{\bf (a)} Correlation between the composition fields of the 2 leaflets (registration) through time. {\bf (ab} Correlation between the  composition field of the bilayer and the local thickness through time. DPPC-DIPC-chol (30:58:12) mixture with percentage of GM1 in the upper leaflet varying from 0 (orange), to 5 (red) and to 10 (purple).}
\label{fig:covephi}
\end{figure*}

We measure the composition correlation between the two leaflets (registration) at a given time step following
\beq
g(\phi^0,\phi^1)=\frac{1}{N} \sum_{i,j} \frac{\phi^0_{ij} \phi^1_{ij} - \langle \phi^0 \rangle \langle \phi^1 \rangle }{s(\phi^0)s(\phi^1)}
\eeq
with $\phi^0$ and $\phi^1$ being the binarized compositions in the upper and lower leaflet respectively, and $N$ the number of sites of the mesh taken into account (here $15^2=225$). We normalize it by the standard deviations $s$ so to get a value ranging from -1 to 1. We check in Fig.~\ref{fig:covephi} (left) that the leaflet compositions are indeed in register with $g(\phi^0,\phi^1)\simeq0.8$.

Similarly, we compute the correlation between the local bilayer thickness $h$ and the local composition $\phi$ of the bilayer (average box composition of both leaflets) defined as
\beq
g(h,\phi)=\frac{1}{N} \sum_{i,j} \frac{h_{ij} \phi_{ij} - \langle h \rangle \langle \phi \rangle }{s(h)s(\phi)}
\eeq
We notice in Fig.~\ref{fig:covephi} (right) a strong correlation with $g(h,\phi)\simeq 0.9$, as expected.

\begin{figure}
\centering
\includegraphics[height=6cm]{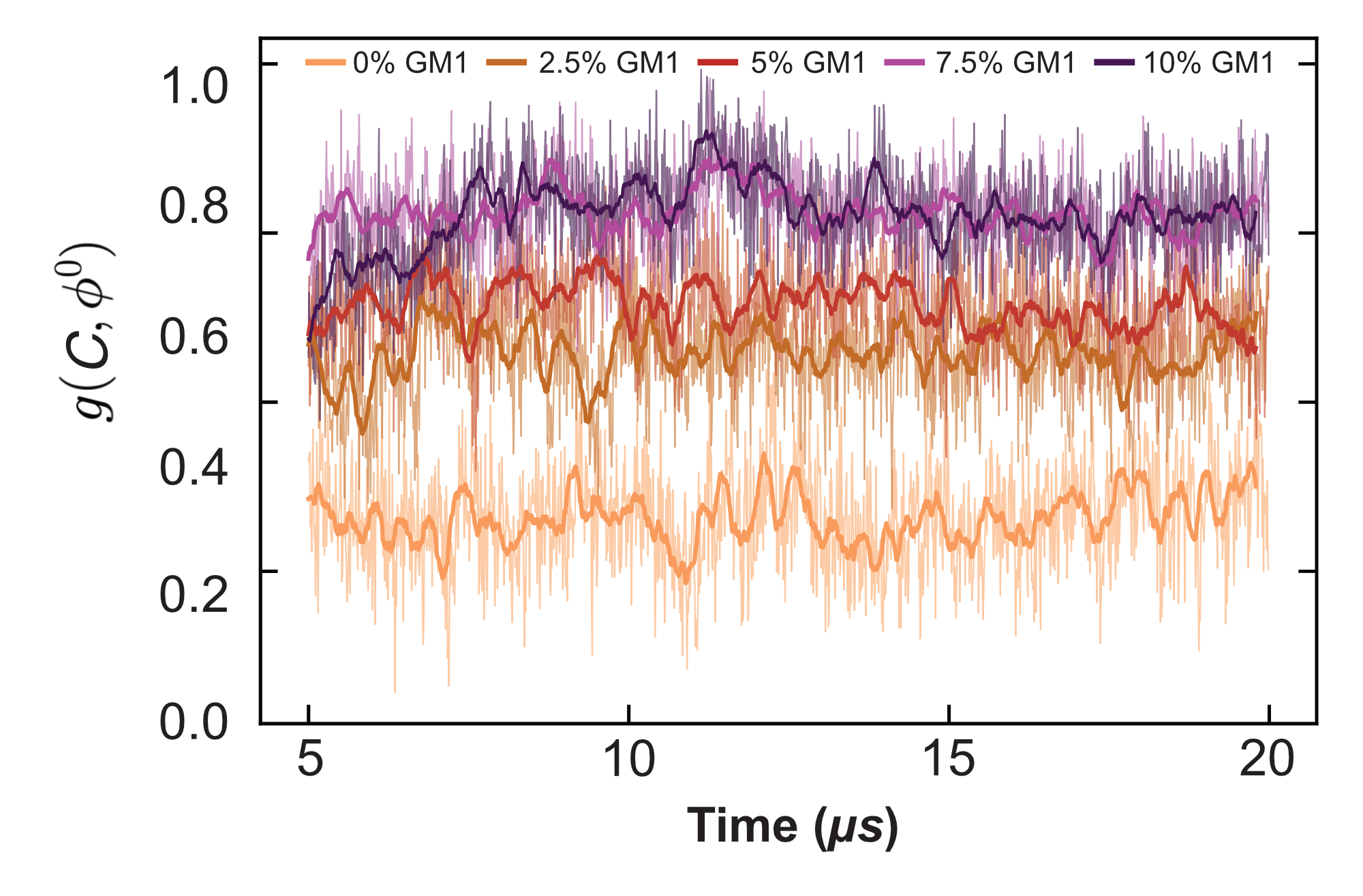} 
\caption{{Correlation between the  composition field of the upper leaflet and the local curvature through time. DPPC-DIPC-chol (30:58:12) mixture with percentage of GM1 in the upper leaflet varying from 0 (orange) to 10 (purple) by steps of 2.5.}}
\label{fig:covC}
\end{figure}

We further compute $g(C,\phi^0)$ providing information about the correlation between the local curvature and the composition of the upper leaflet:
\beq
g(C,\phi^0)=\frac{1}{N} \sum_{i,j} \frac{C_{ij} \phi^0_{ij} - \langle C \rangle \langle \phi^0 \rangle }{s(C)s(\phi^0)}
\eeq
with $C$ being the local upper leaflet curvature. We measure in Fig.~\ref{fig:covC} a high correlation which appears to be higher in the Lo phase where GM1 are inserted. This correlation increases with GM1 fraction.

\section{Using the relations between molecular and mesoscopic parameters}

As an illustrative example, suppose that the Mesoscale simulations are run with an Ising parameter $J_I=0.5k_{\rm B}T$. The line tension measured for mixtures with 10\% of GM1 is $\lambda \simeq 1.2$~pN, so that  $a\simeq5.6$~nm owing to Eq.~(11) of the main text. One can then calculate the corresponding radius associated to the $N$-vertices vesicle simulated at the mesoscale, $R\simeq75$~nm (for $N=2562$, see Eq.~(1)). For a measured local curvature of $\Cpl\simeq0.05$~nm$^{-1}$ at $10$\% GM1 in the outer leaflet, one gets $R\Cpl \simeq 3.8$ and thus $R\Cvesi \simeq 5.8$ as input value for the Lo-phase dimensionless spontaneous curvature in the Mesoscale simulations. 

Conversely one could start from the value of the dimensionless $R\Cvesi$ used in Mesoscale simulations and infer the ensuing real-unit curvature $\Cpl$ and thus the GM1 concentration required to have this value of $R\Cvesi$ in the Mesoscale model.

\section{Relaxation times of boundary fluctuation modes}

Relaxation times of the boundary fluctuations are defined as follows. We consider a quasi-circular domain of average radius $R_0$, the boundary of which fluctuates under thermal agitation. The wavelength associated with the Fourier mode $n$ is $\Lambda_n = 2 \pi R_0/n$. If the domain radius in polar coordinates is $r(\theta,t)=R_0[1+u(\theta,t)]$, we denote by $u_n$ the amplitude of mode $n$. It is a Gaussian random variable of zero mean~[\ref{Esposito2007}]. Its auto-correlation function is defined as $C_n(s) = \langle u_n(t)u_n(t+s) \rangle$, averaged over realization or over a long trajectory. As discussed for example in Ref.~[\ref{Camley2010A}], $C_n(s)$ decays exponentially with $s$, with a characteristic timescale $\tau_n$, also denoted by $\tau(\Lambda_n)$, that we call the relaxation time of mode $n$ in the present work.

Refs.~[\ref{Camley2010A}] and [\ref{Camley2010B}] have discussed in detail the expected dependence of $\tau$ on $n$ in the MD context. Hydrodynamic interactions play a key role because forces locally applied on the domain boundary by the line tension are propagated on the whole boundary by hydrodynamic flows. In the frame of the Saffman-Delbruck theory, a typical lengthscale plays an important role, the so-called Saffman-Delbruck length $L_{\rm SD}=h\eta_m/(2\eta_f)$. Here $L_{\rm SD}$, $\eta_m$ and $\eta_f$ are the viscosities of the membrane and the fluid, respectively, and $h$ is the membrane thickness. In real lipid membranes $L_{\rm SD}$ falls typically between 100~nm and 10~$\mu$m. Depending on the ratio between $L_{\rm SD}$ and $\Lambda_n$, $\tau$ scales differently with $\Lambda_n$. 

On lengh-scales $\Lambda_n \ll L_{\rm SD}$, 2D hydrodynamics inside the membrane dominates and the solvent friction can be neglected, so that
\begin{equation}
\tau_n \simeq \frac{4 h \eta_m}{\lambda} \frac{R_0}{n} \propto \Lambda_n.
\label{tau:2D}
\end{equation}
where $\lambda$ still denotes the line tension at the domain boundary. Note that in this case, if the viscosity of the Lo and Ld phases differ, it can be proven that the value of $ h \eta_m$ entering Eq.~\eqref{tau:2D} is in fact the mean $h \eta_m=(h_{\rm Lo} \eta_{\rm Lo}+h_{\rm Ld} \eta_{\rm Ld})/2$~[\ref{Camley2010A}]. To our knowledge, the viscosity of DLiPC bilayers has not been measured experimentally so far. However, we reasonably assume that Ld phase is much less viscous than the Lo one, so that $h \eta_m \simeq h_{\rm Lo} \eta_{\rm Lo}/2$. Literature where viscosities of lipid mixtures are measured are relatively sparse. Owing to the latter relation, we can extract reliable values of  $h \eta_m$ from experiments on lipid mixtures where the Lo phase is comparable to ours, at comparable temperature. In Ref.~[\ref{Petrov2008}], a 1:2 DOPC (C18:1 dioleoyl PC)/DPPC + 30\% Chol, i.e. 51:25:23 DPPC-DOPC-Chol  is studied, with a Lo phase comparable to ours. A value of $h \eta_m \simeq 4\times10^{-10}$~Pa.m.s is found at about 305~K. We use this value in the main text. 
In addition, $\eta_{\rm f}=0.69$~mPa.s for water at  310~K, hence $L_{\rm SD}\simeq300$~nm, larger than the MD system size. Note that the measured values of membrane viscosities somewhat depends on the experimental technique, so this value is only indicative. 

On lengh-scales $\Lambda_n \gg L_{\rm SD}$, 3D hydrodynamics in the surrounding solvent would dominate, and 
\begin{equation}
\tau_n \simeq \frac{2 \pi \eta_f}{\lambda} \frac{R_0^2}{n^2} \propto \Lambda^2.
\label{tau:3D}
\end{equation}
In the intermediate regime where $\Lambda$ and $L_{\rm SD}$ are comparable, a more complex integral relation has also been derived~[\ref{Camley2010A}].

To measure relaxation times in MD simulations, we use the same discretized dartboard as described in the main text Methods to get the discrete values of $u(\theta,t)$. We then make use of a combination of Fast Fourier Transform (FFT) and Wiener-Khintchine's theorem to compute $C_n(s)$, and we finally extract $\tau_n$ by fitting $\ln [C_n(s)]$ with an affine function. 

In Table~\ref{relax.times}, we give the relaxation times measured this way (in absence of GM1). The slowest mode is on the $\mu$s time-scale for a 12~$\mu$s run (after 8~$\mu$s of equilibration), thus sampling of this mode is relatively poor and the uncertainty on $\tau_2$ is expected to be relatively important. We nevertheless can fit the measured values with a power-law, and we find that $\tau \propto 1/n^{2.5}\propto \Lambda^{2.5}$. In spite of the uncertainties, this suggests that the system size is above the Saffman-Delbruck length $L_{\rm SD}$ where $\tau \propto \Lambda^2$, whereas we would have expected that $ \Lambda \ll L_{\rm SD}$ with known experimental viscosities. This apparent discrepancy either comes from insufficient sampling or from a smaller CG bilayer viscosity. Indeed, it is known to be smaller than the real one by several orders of magnitude~[\ref{denOtter2007},\ref{Vogele2019}], which significantly lowers the value of $L_{\rm SD}$ in CG simulations using the MARTINI force field. If timescales are to be directly extracted from simulations rather than experiments, one must take great care of this delicate issue. 

\begin{table}
\begin{center}
\begin{tabular}{lccc}
\hline
$n$ & 2 & 3 & 4 \\
\hline
$\tau_n$ ($\mu$s) & 0.67 & 0.175 & 0.125 \\
\hline
\end{tabular}
\end{center}
\caption{Relaxation times of domain-boundary fluctuation modes for the MD simulations of the DPPC-DIPC-chol (30:58:12) mixture. 
\label{relax.times}}
\end{table}

\section{Backmapping from the Mesoscale model to the CG one}
\begin{figure}[!h]
\centering
\includegraphics[height=5cm]{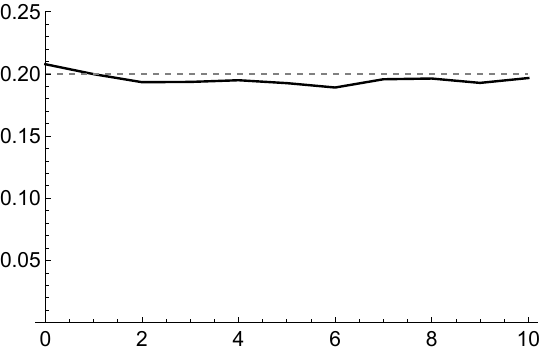} 
\caption{Evolution with time of the fraction of Lo triangles (full line) and expected value of 20\% (dashed line).}
\label{fig:FractionRed}
\end{figure}

In complement of Fig.~12 in the main text, Fig.~\ref{fig:FractionRed} illustrates that the 
 fraction of Lo triangles remains stable during the CG-MD simulation, very close to the initial 20\% fraction before backmapping. 
 
\bigskip

\section*{References}

\small

\begin{enumerate}
\item \label{Seifert1997} U. Seifert, Advances in Physics 46, 13 (1997). \vspace{-1mm}
\item \label{Mouritsen2005} O. Mouritsen, Chemistry and Physics of Lipids 135, 105 (2005).  \vspace{-1mm}
\item \label{Hossein2020}  A. Hossein and M. Deserno, Biophysical Journal 118, 624 (2020).  \vspace{-1mm}
\item \label{Fowler2016} P. W. Fowler, J. Helie, A. Duncan, M. Chavent, H. Koldsø, and M. S. P. Sansom, Soft Matter 12, 7792 (2016).   \vspace{-1mm}
\item \label{Doktorova2017} M. Doktorova, D. Harries, and G. Khelashvili, Phys. Chem. Chem. Phys. 19, 16806 (2017).  \vspace{-1mm}
\item \label{Fernandez1994} L. Fernandez-Puente, I. Bivas, M. D. Mitov, and P. Meleard, Europhysics Letters (EPL) 28, 181 (1994).  \vspace{-1mm}
\item \label{Oradd2009} G. Oradd, V. Shahedi, and G. Lindblom, Biochimica et Biophysica Acta (BBA) - Biomembranes 1788, 1762 (2009).  \vspace{-1mm}
\item \label{Rawicz2000} W. Rawicz, K. Olbrich, T. McIntosh, D. Needham, and E. Evans, Biophysical Journal 79, 328 (2000).  \vspace{-1mm}
\item \label{Landau1986} L. Landau, E. Lifshitz, A. Kosevich, J. Sykes, L. Pitaevskii, and W. Reid, Theory of Elasticity: Volume 7, Course of
theoretical physics (Elsevier Science, 1986).  \vspace{-1mm}
\item \label{Esposito2007} C. Esposito, A. Tian, S. Melamed, C. Johnson, S.-Y. Tee, and T. Baumgart, Biophysical Journal 93, 3169 (2007).  \vspace{-1mm}
\item \label{Camley2010A} B. Camley, C. Esposito, T. Baumgart, and F. Brown, Biophysical Journal 99, L44 (2010).  \vspace{-1mm}
\item \label{Camley2010B} B. Camley and F. Brown, Physical Review Letters 105, 148102 (2010).  \vspace{-1mm}
\item \label{Petrov2008} E. Petrov and P. Schwille, Biophysical Journal 94, L41 (2008).  \vspace{-1mm}
\item \label{denOtter2007} W. den Otter and S. Shkulipa, Biophysical Journal 93, 423 (2007).  \vspace{-1mm}
\item \label{Vogele2019} M. Vogele, J. Kofinger, and G. Hummer, The Journal of Physical Chemistry B 123, 5099–5106 (2019).
\end{enumerate}
\end{document}